\documentclass[fleqn,usenatbib]{mnras}

\usepackage{newtxtext,newtxmath}

\usepackage[T1]{fontenc}

\DeclareRobustCommand{\VAN}[3]{#2}
\let\VANthebibliography\thebibliography
\def\thebibliography{\DeclareRobustCommand{\VAN}[3]{##3}\VANthebibliography}

\usepackage{graphicx}	
\usepackage{amsmath}	
\usepackage{float} 
\restylefloat{table} 
\usepackage{tablefootnote} 
\usepackage{subcaption}
\usepackage{float} 
\usepackage{multirow} 
\usepackage{xurl}
\usepackage{afterpage,refcount}

\restylefloat{table} 
\newcommand{\setfootnotemark}{%
  \refstepcounter{footnote}%
  \footnotemark[\value{footnote}]}

\title[Detecting strong lensing in the NIR]{On the detectability of strong lensing in near-infrared surveys}

\author[P. Holloway et al.]{
Philip Holloway$^{1}$\thanks{E-mail: philip.holloway@physics.ox.ac.uk},
Aprajita Verma$^{1}$,
Philip J. Marshall$^{2,3}$,
Anupreeta More$^{4,5}$,
Matthias Tecza$^1$
\\
$^{1}$Sub-department of Astrophysics, University of Oxford, Denys Wilkinson Building, Keble Road, Oxford OX1 3RH, UK\\
$^{2}$ Kavli Institute for Particle Astrophysics and Cosmology, Department of Physics, Stanford University, Stanford, CA 94305, USA\\
$^{3}$SLAC National Accelerator Laboratory, Menlo Park, CA 94025, USA\\
$^{4}$The Inter-University Centre for Astronomy and Astrophysics (IUCAA), Post Bag 4, Ganeshkhind, Pune 411007, India\\
$^{5}$Kavli Institute for the Physics and Mathematics of the Universe (IPMU), 5-1-5 Kashiwanoha, Kashiwa-shi, Chiba 277-8583, Japan
}

\date{Accepted XXX. Received YYY; in original form ZZZ}

\pubyear{2022}

\begin{document}
\label{firstpage}
\pagerange{\pageref{firstpage}--\pageref{lastpage}}
\maketitle

\begin{abstract}
We present new lensing frequency estimates for existing and forthcoming deep near-infrared surveys, including those from \emph{JWST} and VISTA. The estimates are based on the JAdes extragalactic Ultradeep Artificial Realisations (JAGUAR) galaxy catalogue accounting for the full photometry and morphologies for each galaxy. Due to the limited area of the JAGUAR simulations, they are less suited to wide-area surveys, however we also present extrapolations to the surveys carried out by \emph{Euclid} and the \emph{Nancy Grace Roman Space Telescope}. The methodology does not make assumptions on the nature of the lens itself and probes a wide range of lens masses. The lenses and sources are selected from the same catalogue and extend the analysis from the visible bands into the near-infrared. After generating realistic simulated lensed sources and selecting those that are detectable with SNR>20, we verify the lensing frequency expectations against published lens samples selected in the visible, finding them to be broadly consistent. We find that \emph{JWST} could yield $\sim65$ lensed systems in COSMOS-Web, of which $\sim 25$ per cent have source redshifts $>$4. Deeper, narrower programs (e.g. JADES-Medium) will probe more typical source galaxies (in flux and mass) but will find fewer systems ($\sim25$). Of the surveys we investigate, we find 55--80 per cent have detectable multiple imaging. Forthcoming NIR surveys will likely reveal new and diverse strong lens systems including lensed sources that are at higher redshift (\emph{JWST}) and dustier, more massive and older (\emph{Euclid} NISP) than those typically detected in the corresponding visible surveys.
\end{abstract}




\section{Introduction}\label{Introduction}
Gravitational lensing is the deflection of light emanating from a background source due to the mass of a foreground lens. Strong gravitational lensing occurs when the deflection due to the lens is large enough that multiple images are produced. In galaxy-galaxy lensing this can produce a ring or multiple images of the background source galaxy around the lensing galaxy. Strong lens systems are versatile probes of a wide range of astrophysical and cosmological science cases. For example, lensing can magnify faint, more distant sources to above a telescope's magnitude limit to probe the early universe \citep{Zheng2012,Mcgreer2018,Fan2019}. They can be used to constrain cosmological parameters such as Hubble's Constant  (e.g. \citet{Suyu2013}), the matter distribution within the lens galaxy \citep{Barnabe2012,Treu2002}, and the equation of state (EoS) for dark energy (e.g. \citet{Collett2014}). The magnifying power of strong lensing can also make significant improvements to the reconstruction of the source galaxy dynamics such as described in \citet{Chirivi2020} and \citet{Young2022}. However, there are limited subsamples of lenses suitable to address these science cases since galaxy-galaxy strong lenses remain intrinsically rare with only $\mathcal{O}(10^3)$ found to date \citep{Sonnenfeld2020,Stein2022,ODonnell2022}.
Furthermore, many of these have not been spectroscopically confirmed and span a wide range of confidence levels.\\
Galaxy-scale strong lensing relies on the chance alignment of a background galaxy or quasar with a foreground lens galaxy and so the vast majority of (untargeted) survey images will not contain a strongly lensed galaxy-galaxy or galaxy-QSO system. Selecting rare strong lenses ($\sim1$ in $10^3$ galaxies for HST depth, \citet{Marshall2009}) from galaxy surveys remains a challenge owing to high rates of false positives and the complex morphologies of galaxy and group-scale lenses. Many different methods have been used to find them; these can be categorised into 1) machine learning (e.g. CNN's, \citep{Jacobs2019,He2020}) 2) algorithmic fitting (e.g. YATTALENS, \citet{Sonnenfeld2018}), 3) visual search by researchers \citep{Pawase2014,Moustakas2007} and 4) citizen science \citep{Geach2015,More2016,Marshall2016,Sonnenfeld2020}. Given the forthcoming wide-area surveys (taken with the Vera C. Rubin Observatory, \emph{Euclid} and the \emph{Nancy Grace Roman Space Telescope}), and existing space based telescopes (\emph{HST}, \emph{JWST}), all of these methods will be required to realise the full potential of strong lensing science. 
The narrow surveys of \emph{JWST} will allow a small number of lensed systems and lensing clusters to be analysed in detail (e.g. studying the SMACS 0723 cluster \citep{Caminha2022,Pascale2022}), while the wide-area surveys will allow lens populations to be treated in a statistical fashion, for example to infer the galaxy and dark matter properties \citep{Sonnenfeld2021}, and to infer cosmological parameters such as the Hubble constant, $H_0$ \citep{Sonnenfeld2021b,Park2021}. \\
Historically, lens searches have been conducted in sensitive wide-area surveys utilising the visible, sub-millimetre and radio wavelengths, for example HSC (visible, e.g. \citet{Sonnenfeld2020}), the Herschel Astrophysical Terahertz Large Area Survey (H-ATLAS, sub-mm, \citep{Negrello2010}) and the Cosmic Lens All-Sky Survey (CLASS, radio, \citep{Myers2003,Browne2003}). The detection strategy for strong lenses changes with wavelength; searches in the visible typically select low-redshift early-type galaxy lenses, whereas in far-infrared (FIR) and sub-mm surveys they can be identified by a flux-density selection for the brightest galaxies, where lensed galaxies are much more numerous than their unlensed counterparts \citep{Negrello2010}. 
The dramatic improvement in near-infrared (NIR) detectors over the last two decades has paved the way for wide-area NIR sensitive surveys e.g. with the United Kingdom Infrared Telescope (UKIRT) and the Visible and Infrared Survey Telescope for Astronomy (VISTA). Here we explore the strong lens discovery prospects in the NIR with simulations that are well suited to medium-deep VISTA surveys and those conducted by \emph{JWST}. We also present illustrative figures for \emph{Euclid} and \emph{Roman} surveys (Section \ref{Extrapolating to wide-area Surveys}). 
Predictions for strong lens detection have previously been made by \citet{Collett2015} who focused on visible surveys and used distinct lens and source populations, the former restricted to elliptical galaxies. While this is a sensible choice given that most lenses will be massive ellipticals, here we present estimated lens frequencies based on a non-restrictive prior population and extend our estimates to the NIR. These simulations could include less common strong lenses (e.g. the bulges of late-type spirals, \citet{Treu2011}) and would allow us to understand the complex lensing selection function for an untargeted search in a given survey \citep{Sonnenfeld2022}. In principle, any foreground galaxy can lens a background source and so we draw lenses and sources from the same generated catalogue, allowing detectability and strong-lensing criteria to determine which lens-source pairs would be identified in a given survey. This study presents a framework to assess the detectability of strong lenses for a chosen survey using self-consistent SED and morphological information.\\
The paper is structured as follows. In Section \ref{Data} we detail the data we used to generate our strong-lensing frequency estimates by the method described in Section \ref{Method}. We detail our results in Section \ref{Results} including verifying our simulations against existing lens searches and detailing the properties of the lens and source population in detectable lensing systems. In Section \ref{Discussion} we discuss prospects for \emph{JWST} and lens searches in the NIR and we present our conclusions in Section \ref{Conclusion}. 
In this paper, we assume a flat $\Lambda$CDM cosmology from the Planck Collaboration \citep{Planck2020} with $H_0 = 67.66$ km s$^{-1}$ Mpc$^{-1}$, $\Omega_\Lambda = 0.6889$ and $\Omega_m$ = 0.3111 and we use AB magnitudes throughout.

\section{Data}\label{Data}
\begin{figure}
\centering
\includegraphics[width=\columnwidth]{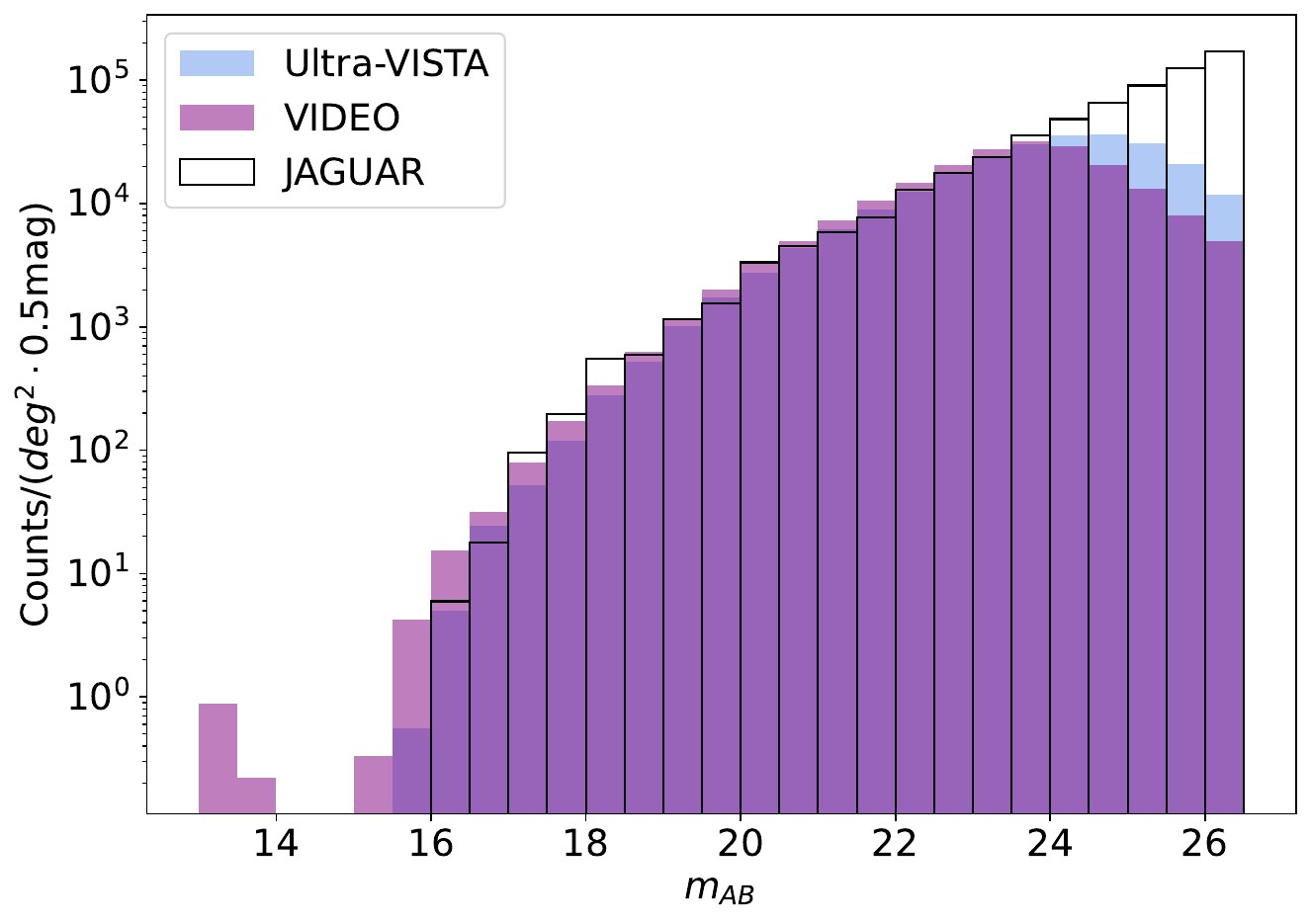}
\caption{Plot of the differential H-band number counts in the NIR surveys VIDEO (XMM and CDFS fields) and UltraVISTA compared to the JAGUAR catalogue. The JAGUAR catalogue aligns well with the NIR surveys up to their respective completeness limits but misses some of the most massive galaxies.}
\label{fig:Video Number Counts}
\end{figure}

To build a representative sample of lensing systems, we first select a suitable galaxy catalogue. We use a single catalogue for both the lens and source galaxies, rather than merging two distinct catalogues. This enables any galaxy to act as a lens of any other, freeing the predictions from biases to lens and source selection, and ensures the corresponding galaxy properties (masses, magnitudes, redshifts etc.) are all self-consistent. The assembled catalogue is required to reach sufficiently high stellar masses (the lensing cross-section increases with lens mass), span a wide redshift range (lenses selected in the visible are typically found at $z\sim$0--1 \citep{Auger2009,Shu2022} with sources at higher redshifts) and reach suitably faint magnitudes. Ideally the magnitude limit of the catalogue would surpass the depth of the target survey since lensing can magnify an otherwise undetectable source galaxy above the magnitude limit. Therefore, we select the JAdes extragalactic Ultradeep Artificial Realisations (JAGUAR) catalogue \citep{Williams2018} as the basis catalogue for our simulations. 
JAGUAR is based on empirical rather than simulated data and contains full SED and morphological information (S\'ersic indices, stellar masses, effective radii etc.) for each galaxy. The galaxy number counts are fitted to empirical stellar mass and luminosity functions for both star-forming and quiescent galaxies and so their proportions in our simulations should be representative of the observations.
The catalogue reaches depths of around $\sim 30$mag, spans a redshift range of $0.2<z<15$ and a stellar mass range of $10^{5.9}<M_{*}<10^{11.7}$ and so also suitable for both source and lens populations. It comprises 10 realisations of 11x11 $arcmin^2$ area.
We verify that the observed number-counts in the VIDEO \citep{Jarvis2013} and UltraVISTA \citep{McCracken2012} surveys are well matched to the JAGUAR catalogue (Fig. \ref{fig:Video Number Counts}). The total area of the JAGUAR catalogue ($0.34\,\deg^2$) is relatively small and so it misses some of the highest mass galaxies ($>10^{11.5}M_{\odot}$) which would lead to an underestimate of the number of detectable lenses in larger surveys. This is an important consideration for the results presented in this study which are therefore optimised towards medium-to-deep ($m>25$), small ($\lesssim 10\,\deg^2$) surveys. We discuss application to wider area surveys in Section \ref{Extrapolating to wide-area Surveys}. Our primary focus in this paper however is on smaller area programs such as those in \emph{JWST} for which this catalogue is most suited.\\
One disadvantage of adopting a catalogue founded on empirically derived relations is that it lacks information on the dark matter profiles which would be available from a cosmological simulation. Therefore we augment the physical parameters from JAGUAR with dark matter properties (profiles and masses) from the Deep Realistic Extragalactic Model (DREaM) galaxy catalogue \citep{Drakos2022}, used for simulating the Roman Ultra-Deep Field (see Section \ref{Forming the galaxy catalogue}). This is a $1\,\deg^2$ catalogue with a magnitude limit similar to JAGUAR reaching $z\sim12$ with the dark matter component simulated from GADGET-2 \citep{Springel2005}. 

\subsection{Adaptions to the galaxy catalogue}\label{Forming the galaxy catalogue}
\subsubsection{Modifications to JAGUAR morphological parameters}
Although the JAGUAR catalogue is well suited to our requirements, there are two areas in which we adapt the supplied values to better reflect realistic galaxy properties, namely the effective radii ($R_{\rm eff}$) and S\'ersic indices ($n_S$).\\
The values of $R_{\rm eff}$ given by the JAGUAR catalogue depend on the galaxy type and redshift. The radii of z<4 star-forming galaxies (SFGs) and quiescent galaxies (QGs) galaxies at all redshifts are based on \citet{vanderwel2014}, using 3D-HST+CANDELS  empirical data based at a rest-frame wavelength of 500nm. The values for star-forming galaxies at z>4 are assigned via a $M_{UV}$-size relation from \citet{Shibuya2015} which predominantly used \emph{J-H HST} bands at such redshifts. 
The JAGUAR values for the S\'ersic indices use results derived from \citet{vanderwel2012}, who in turn uses \emph{HST} F160W CANDELS imaging. We wish to ensure that the $R_{\rm eff}$ and $n_S$ values are consistent; the wavelength dependence on the galaxy profile of changing the effective radius and S\'ersic index simultaneously is small \citep{Kelvin2012} so we simply aim to ensure they are evaluated at the same observed wavelength. This is particularly important for lower redshifts, where the lenses are likely located since differences in these parameters would change the lensing cross-section. 
We therefore shift the $R_{\rm eff}$ values to the observed \emph{H} band, using the $R_{\rm eff}$-$\lambda$ relation from \citet{Vulcani2014}, so they are consistent with the JAGUAR S\'ersic indices. \\
Within a given redshift range, the S\'ersic indices in \citet{Williams2018} are randomly assigned, regardless of galaxy type. Since typical lens galaxies are quiescent, we wish to ensure their mass and light profiles reflected their galaxy type. We therefore reallocate the S\'ersic indices within the JAGUAR catalogue, binned by both redshift and galaxy type, where the $n_S$ distribution for quiescent galaxies is chosen to follow the distribution observed by \citet{Gu2020} from 3D-HST+CANDELS data. 

\subsubsection{Inclusion of dark matter}
The halo masses from the DREaM catalogue galaxies are assigned to JAGUAR galaxies via Subhalo Abundance Matching (SHAM). We order JAGUAR galaxies in order of stellar mass and DREaM halos in order of $V_{\rm peak}$ (the maximum circular velocity over the halo's accretion history), since $V_{\rm peak}$ has been shown to be a good predictor of stellar mass \citep{ChavesMontero2016,Contreras2021}, then pair them accordingly. A dark-matter+baryon simulation would provide a more realistic lens population however our current method allows us to draw our sources and lenses from the same population over a large redshift range. 

\section{Method}\label{Method}
 In brief, our method is as follows. We first select the number of close galaxy pairs ($\Delta\theta<5\arcsec$) in a patch of sky (assuming random angular distribution). We then calculate the lensing properties for such galaxy pairs and determine for how many of these the forefront galaxy strongly lenses the background galaxy. Finally, we impose constraints on the detectability of such systems based on the properties of the survey of interest. \\
 We do not explicitly calculate the lensing cross-section for each galaxy, rather we use a numerical approach, allowing any galaxy to act as a lens (determined by realistic morphological information provided in the base catalogue). The lensing cross-section is indirectly accounted for by constraints on the magnification and relative lens-source positions detailed in Section \ref{Detectability Constraints} that determine which strong lenses would be identified in a typical search. We describe our methodology in detail in the following sections.

\subsection{Frequency of galaxy-galaxy conjunctions}
\begin{figure}
\centering
\includegraphics[width=\columnwidth]{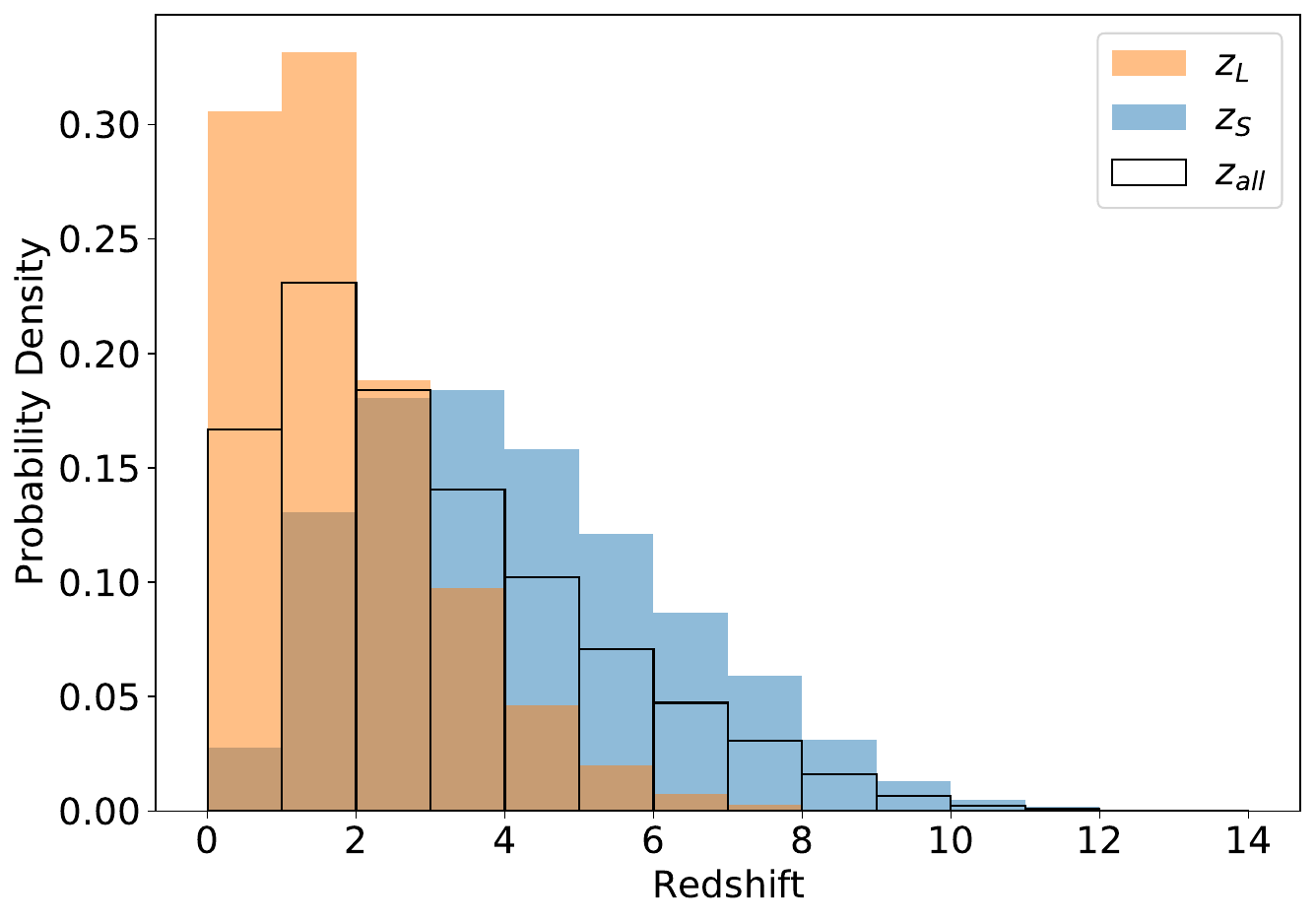}
\caption{Redshift distribution for the original JAGUAR mock catalogue (unfilled), the sample of lenses (orange) and sources (blue). This distribution is before the application of any constraints for detectability or significant lensing and is effectively the prior on the lens and source redshifts.}
\label{fig:redshift_distribution}
\end{figure}

To calculate the number of possible lens systems (neglecting for the moment the Einstein radius of the lens and any detectability and significant magnification criteria, which are applied later) we assume the galaxies are distributed randomly in the sky i.e., we neglect the effect of clustering. For strong lensing to occur the lens and source galaxies must be closely separated on the sky. We choose a $5\arcsec$ upper limit for the galaxy pair separation since this would encompass the vast majority of galaxy-galaxy strong lenses (see fig. 1 in \citet{Collett2015}) while remaining computationally tractable. \\
The derivation for the number of galaxy-galaxy conjunctions is as follows. Consider a set of $N$ objects, distributed randomly within a square of length $L$. We assert that for a source to be lensed it must be located within an angular distance $R$ of the lens. The probability that an object A is within a radius $R$ of a distinct object, B placed randomly within the box is $\frac{\pi R^2}{L^2}$. The mean number of galaxies within radius $R$ of galaxy A is then $\left(N-1\right)\frac{\pi R^2}{L^2}$. Now excluding galaxy A to prevent double counting, the mean number of galaxies close to galaxy B is $\left(N-2\right)\frac{\pi R^2}{L^2}$. By extension, the number of galaxies in proximity to each other is therefore $\frac{\pi R^2}{L^2}\sum_{i=1}^{N-1}i$. Based on the JAGUAR catalogue scaled to $12\,\deg^2$ we use $N = 1.1\times10^8$ objects, with $L = \sqrt{12}\deg$ and choose $R=$5\arcsec as described above. This work was originally motivated by investigating the strong lens population in medium-to-deep NIR surveys with VISTA hence we match our simulated area ($12\deg^2$) to that of the VISTA Deep Extragalactic Observations (VIDEO) Survey \citep{Jarvis2013}. We emphasise here that the $R$ parameter is not the Einstein radius of the lens and has no impact on the lensing potential - it merely limits the number of galaxy pairs to be modelled to reduce computational time; the Einstein radii of the systems are calculated later based on the mass and redshift properties of the lens and source.\\
The above calculation gives $3.1\times10^9$ lens-source pairs for which to determine lensing properties. This simulation is easily scalable to smaller area surveys such as those of \emph{JWST}.
To ensure that a given galaxy does not produce a detectable lens system an unrealistic number of times for the total simulation size, we subdivide our simulation into JAGUAR-sized boxes ($0.34\,\deg^2$) and only count one lensed system from a given lens galaxy from each box; this however only has a minor effect.  To aid computation time, we limit the lens masses to $>10^9M_\odot$ although this makes a negligible difference to the resultant lensing frequencies since typical lenses have much higher mass than this limit.\\
For each galaxy-galaxy conjunction, we draw two redshifts randomly from the redshift population of the JAGUAR catalogue, and assign the lower redshift galaxy (along with its corresponding properties e.g. stellar mass) as the lens, and the higher redshift galaxy as the source. The resultant redshift distribution is shown in Fig. \ref{fig:redshift_distribution}; this is essentially the prior on the lens and source redshifts.

\subsection{Calculating lensing properties}
To calculate lensing properties (Einstein radius, deflection angle etc.) and to generate simulated lensed images, we model each lens galaxy as a circular S\'ersic stellar component with an elliptical Navarro, Frenk and White (NFW, \citep{Navarro1996}) halo profile randomly orientated to our line of sight. Given most of our simulated lens systems have low magnifications (see Fig. \ref{fig:Lens Population}) the overall effect of ellipticity is of order low tens of percent (\citet{Lapi2012}) and thus would not affect our qualitative results. We model the lens and source light with a S\'ersic profile. The S\'ersic profile obeys: 
\begin{equation}
I(R)=I_0\exp\left[-b_n\left(\left(\dfrac{R}{R_e}\right)^{1/n}-1\right)\right]
\end{equation}
where $n$ is the S\'ersic index, $R_e$ is the half-light radius and $b_n$ obeys $\Gamma(2n)$\ = 2$\gamma(2n,b_{n})$ with $\Gamma$ and $\gamma$ the complete and incomplete gamma function respectively. In the case of the lensing galaxy we use the same S\'ersic parameters as the mass profile, i.e. assuming a constant stellar mass-to-light ratio.
An analytic expression for the lensing properties of an elliptical NFW profile is not known so we use a cored steep ellipsoid as a basis function as described in \citet{Oguri2021} to generate such a profile. \\
Using S\'ersic and NFW combined profiles allows a wide variation in lens morphologies and ensures that our simulated galaxies are realistic; we verify below that the properties of our detectable lenses (e.g. Einstein radii and density slopes) agree with well studied lenses. We use the morphological information (i.e. $n_{S}$, $R_{\rm eff}$, $M_*$ etc.) available in the JAGUAR/DReAM catalogues so no further assumptions are required to generate the profiles. Requiring all our galaxies to have an isothermal profile, as is often done, would be unrealistic in the cases where our foreground galaxies are not massive ellipticals and we wish to remain open to lensing from e.g. the bulges of late-type spirals. Furthermore, as demonstrated by \citet{Lapi2012} and \citet{Gavazzi2007}, the difference in lensing cross-section between a de Vaucouleurs+NFW profile compared to an isothermal one is very small and can still provide a good fit to early-type lenses, so we do not expect this to have a material impact on our results. 

\subsection{Detectability constraints}\label{Detectability Constraints}
\begin{figure*}
 \centering
\begin{subfigure}{\textwidth}
  \centering
  \includegraphics[width=\textwidth]{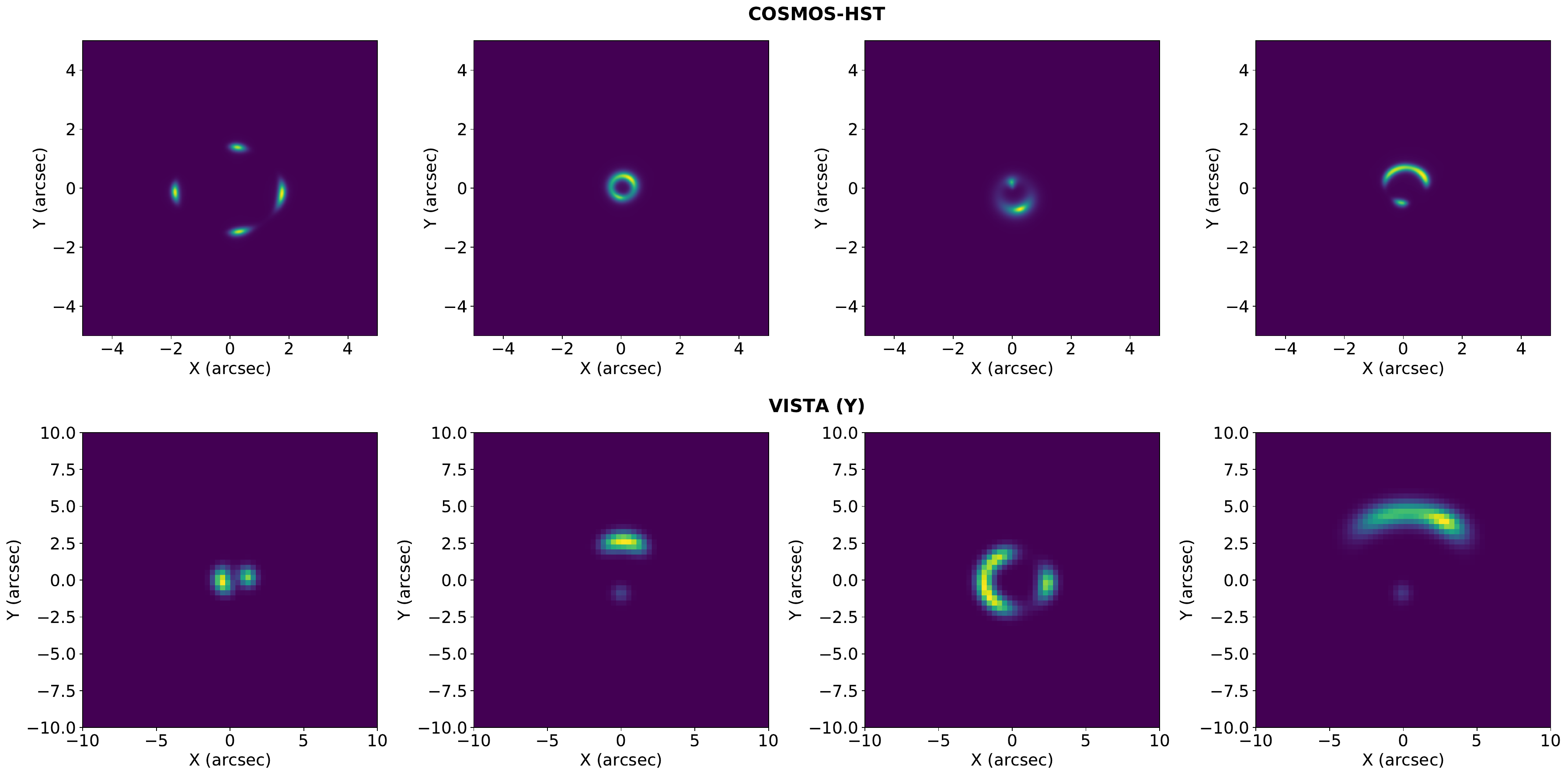}
\end{subfigure}
\caption{Examples of detectable systems in COSMOS-HST and VISTA (Y). In both cases the images are resampled to the survey pixel size and convolved with the relevant PSF.}
\label{Example Detectable Lenses}
\end{figure*}

In order to ascertain which lens systems would be detectable in a given survey we follow the procedure detailed below. Mock images are generated of size $1001\times 1001$ pixels covering a $20$\arcsec $\times 20$\arcsec field. These are then resampled to the pixel scale of each of the surveys of interest and convolved with the appropriate PSF for the survey and bandpass. This resulted in noiseless simulated lensed images for each potential lens. We account for the effect of both the lens galaxy and survey noise in Section \ref{SN_Calculation}. We show example detectable lensing systems from our simulations in Fig. \ref{Example Detectable Lenses}.
To determine whether a given galaxy-galaxy pair would be detectable and classified as a strong lensing system, we apply the following constraints from \citet{Collett2015} to the lensing parameters and generated image: 1) $\mu R_{\rm source}>s$, 2) $\theta_E^2>R_{\rm source}^2+(s/2)^2$, 3) $\mu>3$. Here, $\mu$ is the total magnification, $R_{\rm source}$ is the unlensed source half-light radius, $s$ is the PSF (or seeing, for a ground-based telescope) and $\theta_E$ is the Einstein radius. These constraints correspond to the detection of tangential shear, the ability to resolve the lensed image and requiring significant lensing respectively. We further require the source position to be located within the radial caustic of an NFW+S\'ersic profile, so that multiple images may produced, 4) $\theta_{\rm sep}<\theta_{\rm caustic}$ where $\theta_{\rm sep}$ is the unlensed galaxy separation.\\
We add cuts for resolution, 5) $\theta_E>1.5r_{\rm pixel scale}$ (i.e. 3 pixels in diameter), and depth 6) $m_{\rm lens}<m_{\rm cut}$, (see Table \ref{tab:N_occurence_table_full}), with values tuned to the target survey. Finally, we add a cut for the detectability of the lensed image, 7) $S/N>20$. The signal-to-noise calculation is described in the following section.

\subsubsection{Signal to noise calculation}\label{SN_Calculation}
We calculate the number of photons incident for a given survey, exposure time and filter using the zeropoints in Table \ref{tab:Zeropoints_table}. A zeropoint for the \emph{Roman} telescope was not available so this is calculated in the same manner as \citet{EuclidCollab2022}, using the filter transmission curve provided in the \emph{Roman} documentation\footnote{\url{https://roman.gsfc.nasa.gov/science/WFI_technical.html}}. The total number of photons collected by the telescope due to an object of AB magnitude $m$ is then given by:
\begin{equation}
  n = 10^{\left(\frac{ZP-m}{2.5}\right)}\cdot t_{\rm exp}
\end{equation}
where $t_{\rm exp}$ is the exposure time of the survey. \\
Adding the photon noise from the lens and source galaxies to the background noise of the survey in quadrature, the total noise in a pixel is:
\begin{equation}
N' = \sqrt{\left(\sqrt{n_{\rm lens}}\right)^2 +\left(\sqrt{n_{\rm source}}\right)^2 + \sigma_{b}^2}
\end{equation}
where $n$ denotes the number of photons per pixel from a given source and $\sigma_{b}$ is the background noise from the survey.

\begin{table}
\centering
\begin{center}
\begin{tabular}{p{1.3cm}p{1cm}p{1.3cm}p{0.5cm}} 
\hline
Telescope & Filter & ZP ($1 e^{-} s^{-1}$) & Ref\\
\hline
\hline
\emph{HST} & F814W & 25.95 &\multirow{1}{0.5cm}{\tablefootnote{\url{https://acszeropoints.stsci.edu/}}} \\
\hline
\multirow{4}{1.3cm}{VISTA} & Y & 25.68&\multirow{5}{0.5cm}{\tablefootnote{\citet{Sutherland2015}}}\\
&J & 26.29\\
&H & 26.83\\
&Ks & 26.46\\
\hline
\multirow{8}{1.3cm}{\emph{JWST}}
&F070W & 27.23&\multirow{8}{0.5cm}{\tablefootnote{\url{https://jwst-docs.stsci.edu/jwst-near-infrared-camera/nircam-performance/nircam-absolute-flux-calibration-and-zeropoints}}$^{,}$\tablefootnote{\url{https://jwst-docs.stsci.edu/jwst-near-infrared-camera/nircam-instrumentation/nircam-detector-overview/nircam-detector-performance}}}\\
&F090W & 27.54\\
&F115W & 27.59\\
&F150W & 27.89\\
&F200W & 28.08\\
&F277W & 27.98\\
&F356W & 28.14\\
&F444W & 28.16\\
\hline
\multirow{5}{1.3cm}{\emph{Euclid}}& VIS&$25.50$&\multirow{1}{0.5cm}{\tablefootnote{\citet{Collett2015}}}\\\cline{4-4}
&Y&25.04&\multirow{3}{0.5cm}{\tablefootnote{\citet{EuclidCollab2022}}}\\
&J&25.26\\
&H&25.21\\
\hline
\emph{Roman} & J129& 26.40&\\
\hline
\end{tabular}
  \renewcommand{\thetable}{\arabic{table}}
  \caption{Survey zeropoints used for the lens frequency estimates. A zeropoints was not available for the \emph{Roman} telescope so is calculated here in the same manner as in \citet{EuclidCollab2022} using the filter transmission curves provided in the \emph{Roman} documentation.}
\label{tab:Zeropoints_table}
\end{center}
\end{table}

We calculate the noise background of the survey from available magnitude limits in the literature. Where only point source rather than blank-field depths were available (namely \emph{Euclid} NISP, \emph{Roman} and \emph{JWST}), we apply aperture correction to a radius equal to the 80 per cent encircled energy from the PSF and from this calculate the noise per pixel. This process produces a noise map, tending to the (constant) survey limit towards the image edges, but accounting for the increased noise due to the lensing galaxy in the centre. We do not include noise associated with imperfect lens subtraction. Using our generated lensed system images, we then calculate the signal-to-noise ratio for each pixel in the image. For all regions of connected (i.e. adjacent) pixels in the array with S/N $\geq$ 1, we calculate the total signal to noise according to:
\begin{equation}
S/N = \frac{\sum_i{S_i}}{\sqrt{\sum_i{N_i^{'2}}}}
\end{equation} where we sum over all the connected pixels in a given region. If any region has a S/N value $>$20, the lens system is deemed detectable. In this manner, we account for the difficulty in identifying a source close to very bright lenses, as well as the variation in depth with each survey. By focusing only on the connected pixels, we also account for the fact that collections of adjacent brighter pixels would be easier to identify than if those pixels were dispersed.
Many lens searches \citep{Jacobs2017,Faure2008,More2016} are based on images which are not lens subtracted. This makes source identification more difficult due to blending with the lens galaxy. We estimate the results for searches without lens subtraction by masking out pixels from the source image which are fainter than the lens at that position, then calculating the S/N in the same manner as described above. We can also use our signal to noise calculation to determine the number of systems in which multiple lensed images are detectable, described further in Section \ref{multiple imaging}.

\section{Results}\label{Results}
\subsection{Verifying the simulations}
Here we detail verification of our simulations. We measure the density slopes of detectable lenses in \emph{HST} to ensure that on average, our results tally with the finding of the `bulge-halo conspiracy' - that lensing massive elliptical galaxies have broadly isothermal slopes \citep{Koopmans2009}. We further compare the Einstein radii of SLACS galaxies with those of similar galaxies in our simulated catalogue to ensure our choice of lensing profile does not have undue influence on the resultant lens properties.
\subsubsection{Measurement of the mass-density slope}
The combined stellar and dark matter profile of early-type galaxies (i.e. typical strong lenses) has been found to be roughly isothermal (e.g. \citet{Koopmans2009}), referred to as the `Bulge-halo conspiracy'.
We verify that the resultant surface density gradients, $\gamma_{\rm surf}$ from our composite NFW+S\'ersic profiles are on average isothermal. The measured values of the surface density slope depend on the method of measurement. We first define the surface density slope $\gamma_{\rm surf} = - d$log$_{10}(M(R))/d$log$_{10}(R)$ where $M(R)$ is the mass enclosed by a cylindrical radius R. We then choose four methods as follows: 
\begin{itemize}
    \item The mean value of $\gamma_{\rm surf}$ for R values between $10^{-4}\theta_E$ and $\theta_E$ with $\Delta R=10^{-4}\theta_E$,
    \item The best-fit of a straight line in log-log space for R vs M(R), with R values between $10^{-4}\theta_E$ and $\theta_E$ and $\Delta R=10^{-4}\theta_E$,
    \item The best-fit of a straight line in log-log space for R vs M(R), with R values between $10^{-4}\theta_E$ and $\theta_E$ with $R_{i+1}=(1+10^{-4})R_{i}$ (i.e. giving greater weight to lower values),
    \item The weighted mean value of $\gamma_{\rm surf}$, weighted by the enclosed mass in an cylindrical annulus at each radius, for R values between $10^{-4}\theta_E$ and $\theta_E$ with $\Delta R=10^{-4}\theta_E$. This is the 2-dimensional equivalent to the mass-averaged slope given by \citet{Dutton2014}.
\end{itemize} 
The slopes of our simulated lenses are typically marginally shallower than isothermal; this would have the effect of reducing our lensing cross-section (e.g. \citet{Mandelbaum2009}), making our estimates more conservative. The median values for each method (top-bottom) are: $0.822\pm0.179$, $0.926\pm0.206$, $1.077\pm0.349$, and $0.859\pm0.190$. For comparison, an isothermal profile has surface density gradient of $\gamma_{\rm surf}=1$ with which our measurements are consistent within the scatter. \citet{Sonnenfeld2013} finds that the density slope varies with stellar mass density and redshift so we should not expect all our galaxies to be exactly isothermal. \citet{Koopmans2009} identified a scatter of $\sigma_\gamma \leq 0.2$ for SLACS lenses. \citet{Sonnenfeld2013} calculated the power law slope for the SL2S lens sample; the raw scatter in the slope was $0.2$, reduced to an intrinsic scatter of $\sigma_\gamma = 0.12$ after accounting for the evolution in $\gamma$ with mass and redshift. We do not account for dependence on such factors here so our measurements of the scatter in the mass-density slope also agree with those observed.

\subsubsection{Confirming the presence of realistic lensing galaxies}
To verify the adapted JAGUAR catalogue could reproduce observed lenses we search for analogues to the lenses found in the Sloan Lens ACS (SLACS) Survey \citep{Bolton2008} and compare the Einstein radii measurements. The SLACS lenses were analysed by \citet{Grillo2009} and \citet{Etherington2022}. We use the values from \citet{Grillo2009} since their measurements of the enclosed mass did not depend on the effective radius of the galaxy, as was the case in \citet{Etherington2022}, but purely on the Einstein radius.
We identify comparable lensing galaxies to SLACS in our simulated galaxy catalogue (matching the mass and redshift of the SLACS lenses to galaxies in our catalogue to within a factor of 0.9-1.1 in mass and $\pm 0.2$ in lens redshift) and single out the galaxy with the closest stellar-mass fraction to those in \citet{Grillo2009}. 55 out of 57 galaxies in \citet{Grillo2009} can be matched to similar galaxies in our simulation in this way. i.e. most of these lensing galaxies are present within our simulations. The remaining two do not pass these tests due to a lack of sufficiently massive galaxies in our catalogue. For those which pass, we then calculate the corresponding Einstein radii which are in good agreement with the SLACS values from \citet{Grillo2009} with a mean ratio of $0.96 \pm 0.07$.

\subsection{The observable lens population}
\subsubsection{Existing and forthcoming surveys investigated}
We investigate the strong lens populations in a range of ground-based and space-based surveys spanning a broad range of depths and wavelengths. We first compare our results to 3 existing \emph{HST} searches of the COSMOS field \citep{Faure2008,Jackson2008} and archival images (\citet{Pawase2014}). We then focus on surveys for which lens searches have not yet taken place. We compare the strong lensing population in COSMOS-HST to that which might be found using the ground-based UltraVISTA survey (\citet{McCracken2012}), a $1.8\,\deg^2$ which uses the \emph{Y, J, H} and \emph{Ks} bands on the VISTA telescope within the COSMOS field. We also provide estimates for the same bands in the VIDEO survey (\citet{Jarvis2013}), a $12\,\deg^2$ NIR survey of the ELAIS-S1, XMM–Newton and extended Chandra Deep Field-South fields. The UltraVISTA survey is split into `deep' and `ultradeep' stripes, the former being of comparable depth to the VIDEO survey so we just consider the ultradeep regions in this work, which cover half the area. We then investigate strong-lens prospects for \emph{JWST}. The largest (and relatively shallow by \emph{JWST} standards) survey COSMOS-Web (\citet{Casey2022}) has point-source depths $m_{\rm F115W} \sim27$mag, covering an area of $\sim1900\,arcmin^2$. The JADES medium and deep program (\citet{Rieke2019}) by comparison are significantly deeper ($m_{\rm F115W}=29.6-30.6$mag for the medium and deep surveys respectively) but much smaller area ($190$ and $40\,arcmin^2$ respectively). These surveys are chosen since they span a representative range of \emph{JWST} surveys (see \citet{Casey2022} for an overview).  We can provide lensing frequency estimates for any JWST surveys not included here on request. Finally, we comment on expectations for the forthcoming Euclid wide-area Survey \citep{EuclidCollab2022_WIDE_AREA_SURVEY} and Roman High Latitude Wide-Area surveys\footnote{\url{https://roman.gsfc.nasa.gov/high_latitude_wide_area_survey.html}} which are discussed further in Section \ref{Extrapolating to wide-area Surveys}. We do not consider the Legacy Survey of Space and Time (LSST) in this work, since this is a wide area optical survey so is less suited to our simulations.
\begin{figure*}
 \centering
\begin{subfigure}{0.5\textwidth}
  \centering
  \includegraphics[height=170pt]{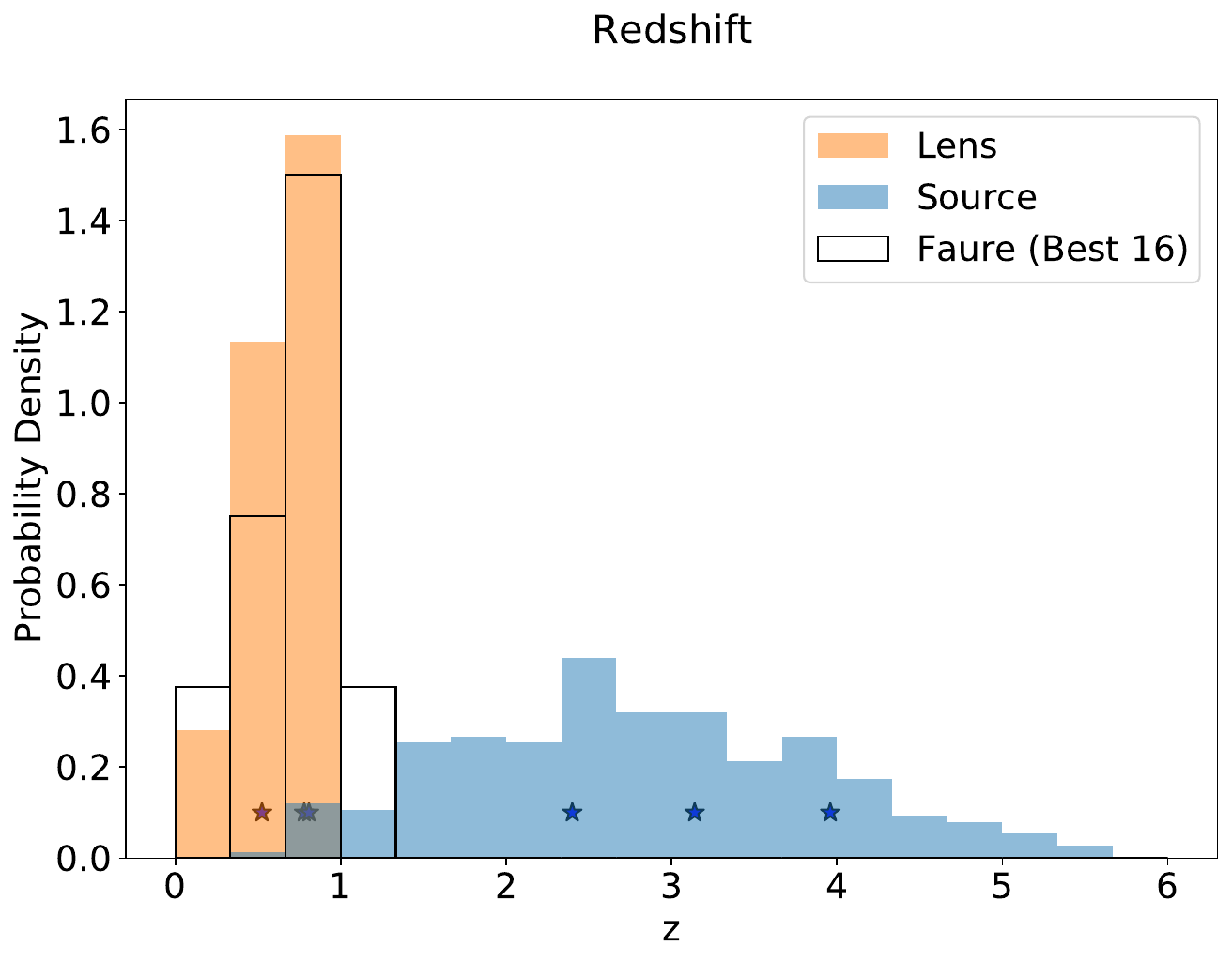}
  \caption{}
\end{subfigure}%
\begin{subfigure}{0.5\textwidth}
  \centering
  \includegraphics[height=170pt]{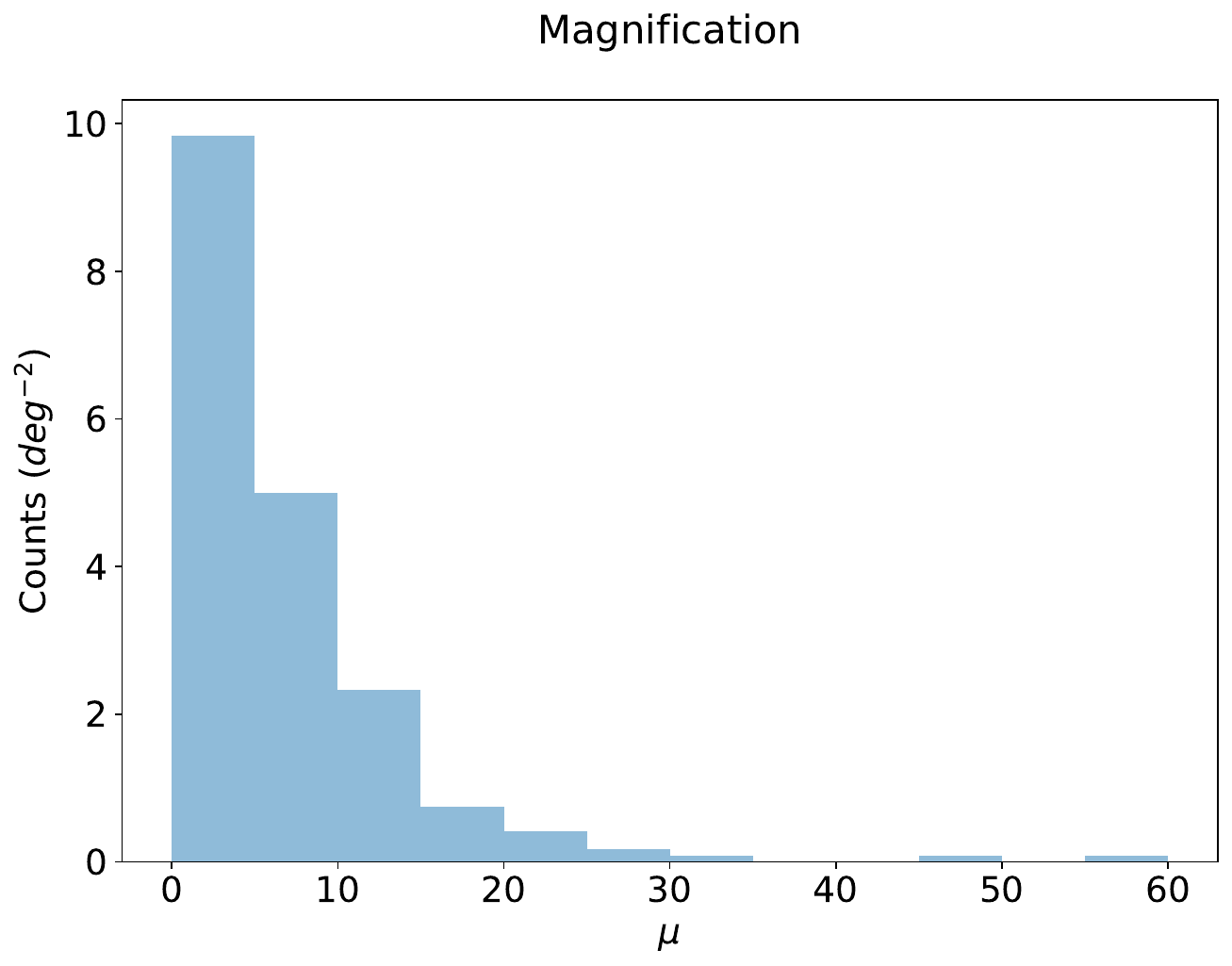}
  \caption{}
\end{subfigure}
\begin{subfigure}{0.5\textwidth}
  \centering
  \includegraphics[height=170pt]{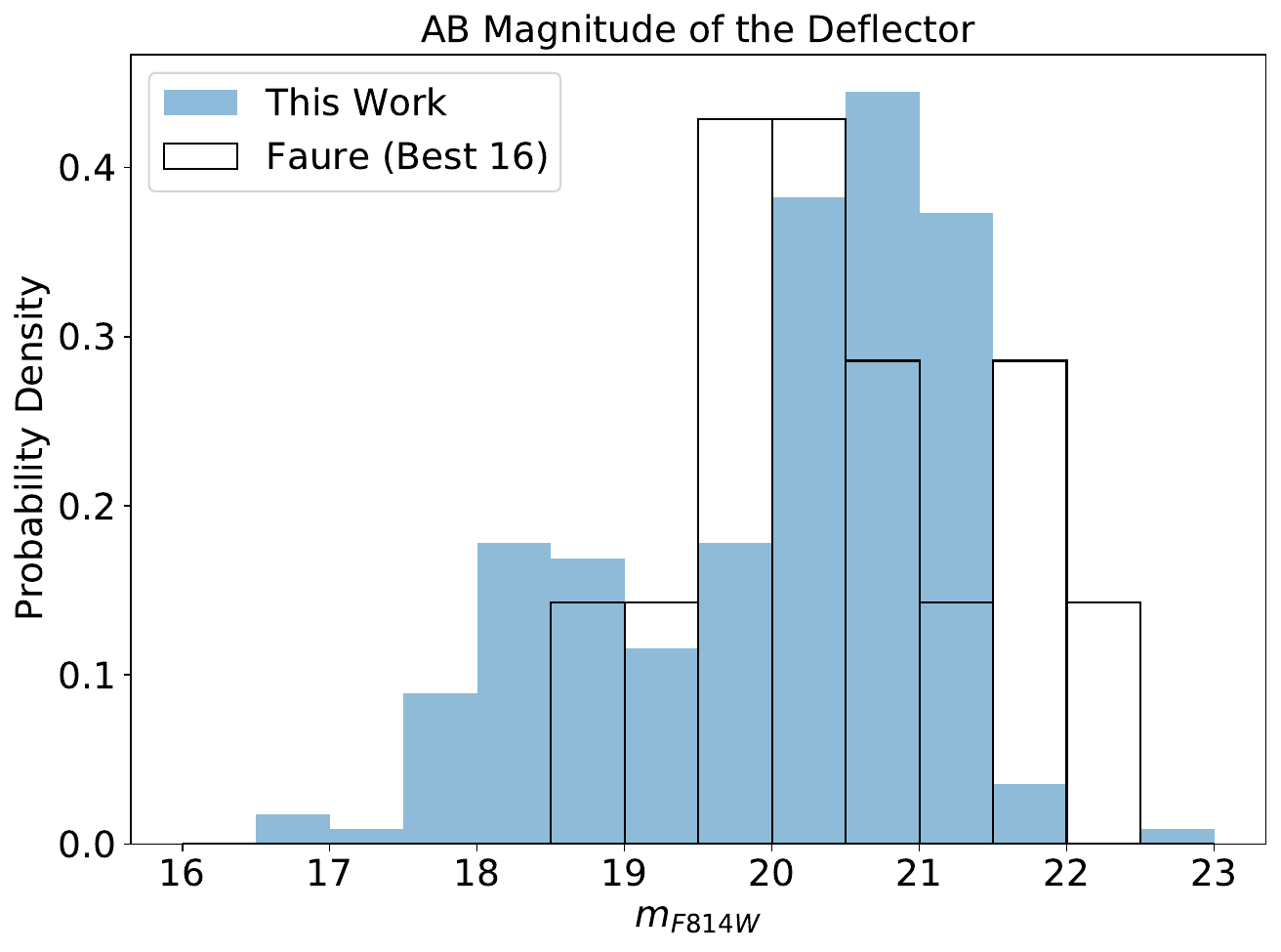}
  \caption{}
\end{subfigure}%
\begin{subfigure}{0.5\textwidth}
  \centering
  \includegraphics[height=170pt]{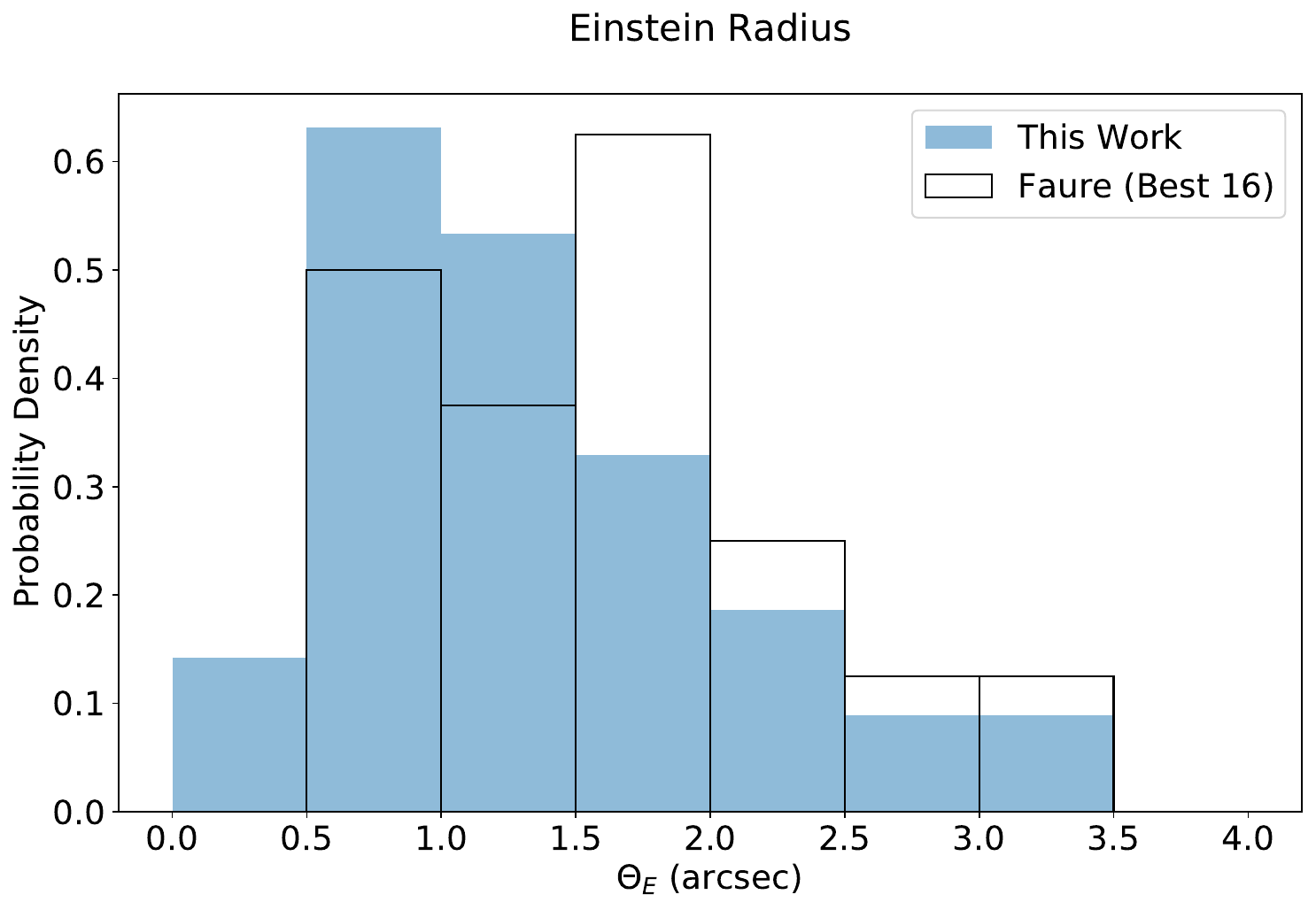}
  \caption{}
\end{subfigure}
\caption{Properties of the detectable lens population at $20\sigma$ a) The simulated and observed lens/source redshift population of lenses detectable in COSMOS. The majority of the observed sources have not been spectroscopically confirmed however we plot redshifts of the sources where known (starred). b) The magnification distribution for the simulated COSMOS lens systems. c) The simulated and observed F814W magnitude of the lenses in COSMOS d) The Einstein radii of simulated and observed lenses in the COSMOS survey.}
\label{fig:Lens Population}
\end{figure*}

\subsubsection{Comparison with existing lens searches}\label{Comparison to existing searches}
The estimated values for the number of strong lenses detectable in the above surveys are shown in Table \ref{tab:N_occurence_table_p1} (verification to previous lens searches), \ref{tab:N_occurence_table_p2} (estimates for smaller survey areas) and \ref{tab:N_occurence_table_p3} (extrapolation to wider area surveys).

\begin{table*}
\begin{subtable}{\textwidth}
\begin{center}
\begin{tabular}{p{1.3cm}p{2cm}p{1.3cm}p{1.5cm}p{1cm}p{1.2cm}p{1.7cm}p{1.2cm}p{1.5cm}p{0.5cm}}
 \hline
 Telescope & Survey & Filter & PSF/Seeing (") & Area ($\deg^2$) & $m_{\rm cut}$ (lens) &\multirow{2}{1.7cm}{ 5$\sigma$ limit ($m_{AB}$ pix$^{-1}$)}&\multirow{2}{1.2cm}{$N$ $(\deg^{-2})$}&\multirow{2}{1.5cm}{$N$ (per survey)}&Ref.\\ [0.5ex] 
\hline
\hline
\emph{HST} & COSMOS (F) & F814W &0.12&1.6&25.0
\multirow{1}{0.5cm}{\setfootnotemark\label{first}}
&30.42&21&34 (31)&
\multirow{1}{0.5cm}{\setfootnotemark\label{second}}\\
& COSMOS (J) & F814W &0.12&1.6&25.0
\multirow{1}{0.5cm}{\setfootnotemark\label{third}}
&30.42&49&80 (75)&
\multirow{1}{0.5cm}{\setfootnotemark\label{forth}}\\
& COSMOS (All) & F814W &0.12&1.6&26.7&30.42&54&88 (82)&
\multirow{1}{0.5cm}{\setfootnotemark\label{fifth}}\\
& Archive & F814W &0.12&6.0&25.4&28.70&17&100 (91)&
\multirow{1}{0.5cm}{\setfootnotemark\label{sixth}}\\
\hline
\end{tabular}
\end{center}
  \caption{Lensing frequency estimates for detectable lenses (at S/N=20, suitable for a visual search) using the HST telescope for comparison to existing lens searches. (F) and (J) refers to the lens searches in the COSMOS field by \citet{Faure2008} and \citet{Jackson2008} respectively, and the archive search refers to that of \citet{Pawase2014}. COSMOS (All) refers to a theoretical lens search of the COSMOS field in which none of the constraints imposed by \citet{Faure2008} or \citet{Jackson2008} were applied, i.e. a untargeted search. The bracketed terms in the penultimate column refers the lensing frequency which might be expected in a lens search using images which have not been lens subtracted.}
\label{tab:N_occurence_table_p1}
\end{subtable}
\\
\begin{subtable}{\textwidth}
\begin{center} 
\begin{tabular}{p{1.3cm}p{2cm}p{1.3cm}p{1.5cm}p{1cm}p{1.2cm}p{1.7cm}p{1.2cm}p{1.5cm}p{0.5cm}}
 \hline
 \hline
\multirow{9}{1.3cm}{VISTA} & \multirow{4}{1.4cm}{VIDEO}&\emph{Y}&0.8&12&25.2&26.97&7.0&84 (53)&
\multirow{4}{0.5cm}{\setfootnotemark\label{seventh}}\\
&&\emph{J}&0.8&12&24.7&26.49&5.4&65 (40)&\\
&&\emph{H}&0.8&12&24.2&25.99&4.2&50 (33)&\\
&&\emph{Ks}&0.8&12&23.8&25.59&3.3&40 (29)&\vspace{0.5cm}\\
&\multirow{4}{1.6cm}{UltraVISTA\\(UD)}&\emph{Y}&0.77&0.9&25.8&27.89&15&13 (8.0)&
\multirow{4}{0.5cm}{\setfootnotemark\label{fifteenth}}\\
&&\emph{J}&0.77&0.9&25.6&27.69&15&14 (7.7)&\\
&&\emph{H}&0.76&0.9&25.2&27.29&14&13 (7.8)&\\
&&\emph{Ks}&0.78&0.9&24.9&26.99&13&11 (7.1)&\\
 \hline
\multirow{20}{1.3cm}{\emph{JWST}}&\multirow{4}{1.4cm}{COSMOS-Web}
&F115W &0.040&0.54&26.8&29.34&32&17 (16)& 
\multirow{4}{0.5cm}{\setfootnotemark\label{eighth}}\\
&&F150W &0.050&0.54&27.1&29.44&47&25 (22)&\\
&&F277W &0.092&0.54&27.5&29.59&110&59 (48)&\\
&&F444W &0.145&0.54&27.4&29.89&110&62 (47)&\vspace{0.5cm}\\

&\multirow{7}{1.4cm}{JADES-Medium}
&F070W &0.029&0.053&28.8&31.53&200&10 (9.2)&
\multirow{7}{0.5cm}{\setfootnotemark\label{ninth}}\\
&&F090W &0.033&0.053&29.4&32.03&340&18 (15)&\\
&&F115W &0.040&0.053&29.6&32.16&400&21 (16)&\\
&&F150W &0.050&0.053&29.7&32.04&390&21 (15)&\\
&&F200W &0.066&0.053&29.8&32.33&460&24 (16)&\\
&&F277W &0.092&0.053&29.4&31.47&400&21 (13)&\\
&&F356W &0.116&0.053&29.4&31.67&400&21 (12)&\\
&&F444W &0.145&0.053&29.1&31.57&340&18 (10)&\vspace{0.5cm}\\

&\multirow{7}{1.4cm}{JADES-Deep}
&F090W &0.033&0.011&30.3&32.93&640&7.2 (5.3)&
\multirow{7}{0.5cm}{\setfootnotemark\label{tenth}}\\
&&F115W &0.040&0.011&30.6&33.16&780&8.6 (5.9)&\\
&&F150W &0.050&0.011&30.7&33.04&740&8.2 (5.0)&\\
&&F200W &0.066&0.011&30.7&33.23&790&8.7 (4.6)&\\
&&F277W &0.092&0.011&30.3&32.37&660&7.3 (3.6)&\\
&&F356W &0.116&0.011&30.2&32.47&620&6.9 (3.2)&\\
&&F444W &0.145&0.011&29.9&32.37&520&5.7 (2.7)&\\
\hline
\end{tabular}
\end{center}
\caption{As above, for forthcoming and existing NIR surveys.}
\label{tab:N_occurence_table_p2}
\end{subtable}

\begin{subtable}{\textwidth}
\begin{center}
\begin{tabular}{p{1.3cm}p{2cm}p{1.3cm}p{1.5cm}p{1cm}p{1.2cm}p{1.7cm}p{1.2cm}p{1.5cm}p{0.5cm}}
 \hline
 \hline
\emph{Euclid} & Wide &\emph{VIS}&0.17&15,000 &26.2&27.41&6.3&95,000 (86,000)&
\multirow{8}{0.5cm}{\setfootnotemark\label{eleventh}}\\
&&\emph{Y}&0.22&15,000 &24.3&25.49&1.8&28,000 (28,000)&\\
&&\emph{J}&0.30&15,000 &24.5&25.69&3.8&58,000 (51,000)&\\
&&\emph{H}&0.36&15,000 &24.4&25.59&4.1&61,000 (55,000)&\\
 \hline
\emph{Roman} &High \mbox{Latitude} Wide-Area&J129&0.1&1,700&27.1&29.12&52&88,000 (72,000)&
\multirow{1}{0.5cm}{\setfootnotemark\label{thirteenth}}\\
 \hline
\end{tabular}
  \caption{As above, for wider area surveys included for comparison to previous predictions. These values represent extrapolations from our dataset as discussed in detail in Section \ref{Extrapolating to wide-area Surveys} .}
\label{tab:N_occurence_table_p3}
\end{center}
\end{subtable}
\caption{}
\label{tab:N_occurence_table_full}
\end{table*}

\footnotetext[\getrefnumber{first}]{To make our results comparable with \citet{Faure2008}, we adopt constraints of $0.2<z_L<1.0$, $M_V<-20$ and $m_{\rm F814W}<25$.}
\footnotetext[\getrefnumber{second}]{\citet{Taniguchi2009}}
\footnotetext[\getrefnumber{third}]{To make our results comparable with \citet{Jackson2008}, we adopt constraints of $\theta_{E}<2.5$\arcsec and $m_{\rm F814}<25$.}
\footnotetext[\getrefnumber{forth}]{\citet{Taniguchi2009}}
\footnotetext[\getrefnumber{fifth}]{\citet{Taniguchi2009}}
\footnotetext[\getrefnumber{sixth}]{\citet{Pawase2014}}
\footnotetext[\getrefnumber{seventh}]{\citet{Varadaraj2023}}
\footnotetext[\getrefnumber{fifteenth}]{\citet{Moneti2023,McCracken2012}}
\footnotetext[\getrefnumber{eighth}]{\citet{Casey2022}}
\footnotetext[\getrefnumber{ninth}]{\citet{Rieke2019}}
\footnotetext[\getrefnumber{tenth}]{\citet{Rieke2019}}
\footnotetext[\getrefnumber{eleventh}]{\citet{EuclidCollab2022_WIDE_AREA_SURVEY}}
\footnotetext[\getrefnumber{thirteenth}]{\url{https://roman.gsfc.nasa.gov/science/ETC2/ExposureTimeCalc.html}, using a PSF-fitted point source and zodiacal light 1.4x the minimum.}
The values in Table \ref{tab:N_occurence_table_p1} show simulated lensing frequencies for previously undertaken lens searches of \emph{HST} imaging, along with those of an untargeted search of the COSMOS field (i.e. with no prior redshift or magnitude cuts applied). Unless explicitly stated, the results of our simulations for a COSMOS-HST search in this work refer to this theoretical search, i.e. without the constraints applied by previous studies. In this current section (\ref{Comparison to existing searches}), we \emph{do} apply such constraints to allow comparison with the results of previous studies. \citet{Faure2008} details a visual search of the $1.64\,\deg^2$ \emph{HST} COSMOS survey. They present 67 candidates, of which 20 display multiple images or large arcs. Our prediction for the COSMOS survey (34 of which 31 are detectable without subtraction) is within the range of identified lenses in \citet{Faure2008}.\\
\citet{Jackson2008} conducted a visual search of all galaxies in COSMOS using \emph{HST} imaging with $i<25$mag and found 2 certain, 1 probable and $\mathcal{O}(100)$ possible (but unlikely) lens systems beyond those of \citet{Faure2008}. Their cutout size limited the detectable Einstein radii to $\theta_E<2.5$\arcsec and they only identified 50 per cent of the best lenses identified by \citet{Faure2008} through their manual search. Loosening our constraints to match those of \citet{Jackson2008} produces 80 lenses from our simulations. Due to the completeness of the manual search compared to the \citet{Faure2008}'s more targeted inspection, it is perhaps unsurprising that our estimates are higher than the number found through this method although they are in line with the original expectations from \citet{Jackson2008} of $\sim 100$.\\
\citet{Pawase2014} also conducted a lens search using all the available archival \emph{HST/ACS} F814W-band imaging data. Together these cover a wider area than COSMOS (6.03$\,\deg^2$) but also have a range of depths; we used an average $5\sigma$ depth of 25.45. They identified 13 A-grade, 18 B-grade and  9 C-grade lens systems (totalling 40).
\citet{Pawase2014} attribute the reduced number density of lenses in this field compared to COSMOS down to cosmic variance and the reduced number of inspectors compared to \citet{Faure2008} which could also reduce the number identified compared to our estimates. Our simulations for this data predict a higher value of 100 lenses (91 without subtraction) which may be attributed to the broad range of depths of the archival images, or indicate more lenses yet to be found.\\
The masses, magnifications, magnitudes and redshifts of the simulated lens and source populations detectable in the COSMOS survey are shown in Fig. \ref{fig:Lens Population} and where possible compared to the observed population from \citet{Faure2008}. The initial COSMOS search did not use lens-subtracted images so in these plots we apply the constraints for a non-lens subtracted search as described in Section \ref{SN_Calculation}; this ensures the images deemed detectable are on the outskirts of the lens galaxy light distribution where they should be more easily identifiable. The Einstein radii agree reasonably with the \citet{Faure2008} population extending to the same maximum radii and a two-sample Kolmogorov--Smirnov test gives a p-value of 0.17 suggesting the null hypothesis that the populations are drawn from the same population cannot be rejected. The simulated lens magnitudes are typically brighter than observed (median magnitude difference $\Delta m \sim 0.2$). This could be due to cosmic variance in a small field and small number statistics for the total number of lenses (we only use the best 16 systems identified by \citet{Faure2008} here to prevent contamination by false positives).

\subsection{Lens and source population properties}
\begin{figure*}
\centering
\begin{subfigure}{0.33\textwidth}
    \centering
    \includegraphics[width=\textwidth]{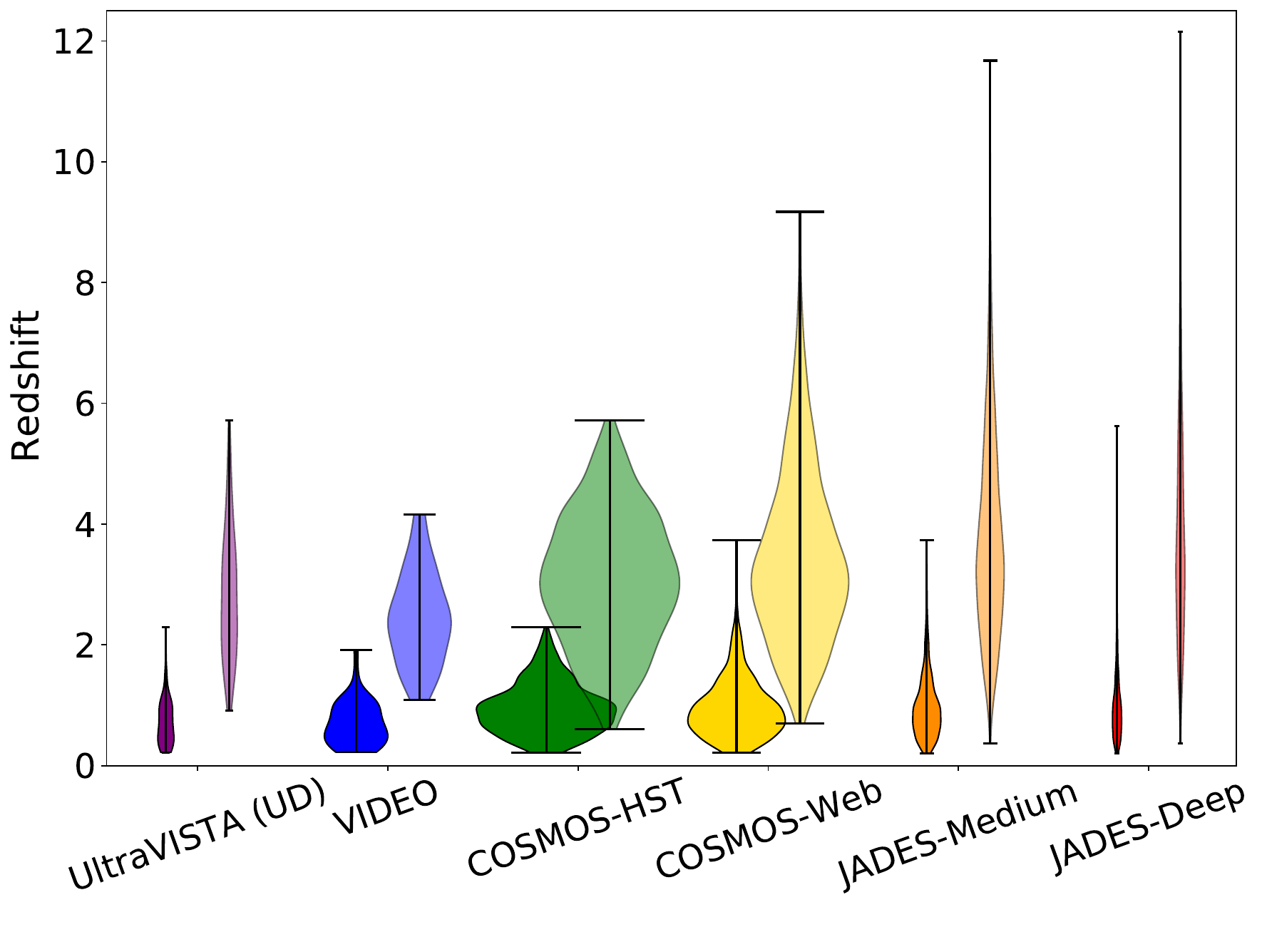}
    \caption{}
    \label{fig:ensemble_boxplots_redshift}
\end{subfigure}
\begin{subfigure}{0.33\textwidth}
    \centering
    \includegraphics[width=\textwidth]{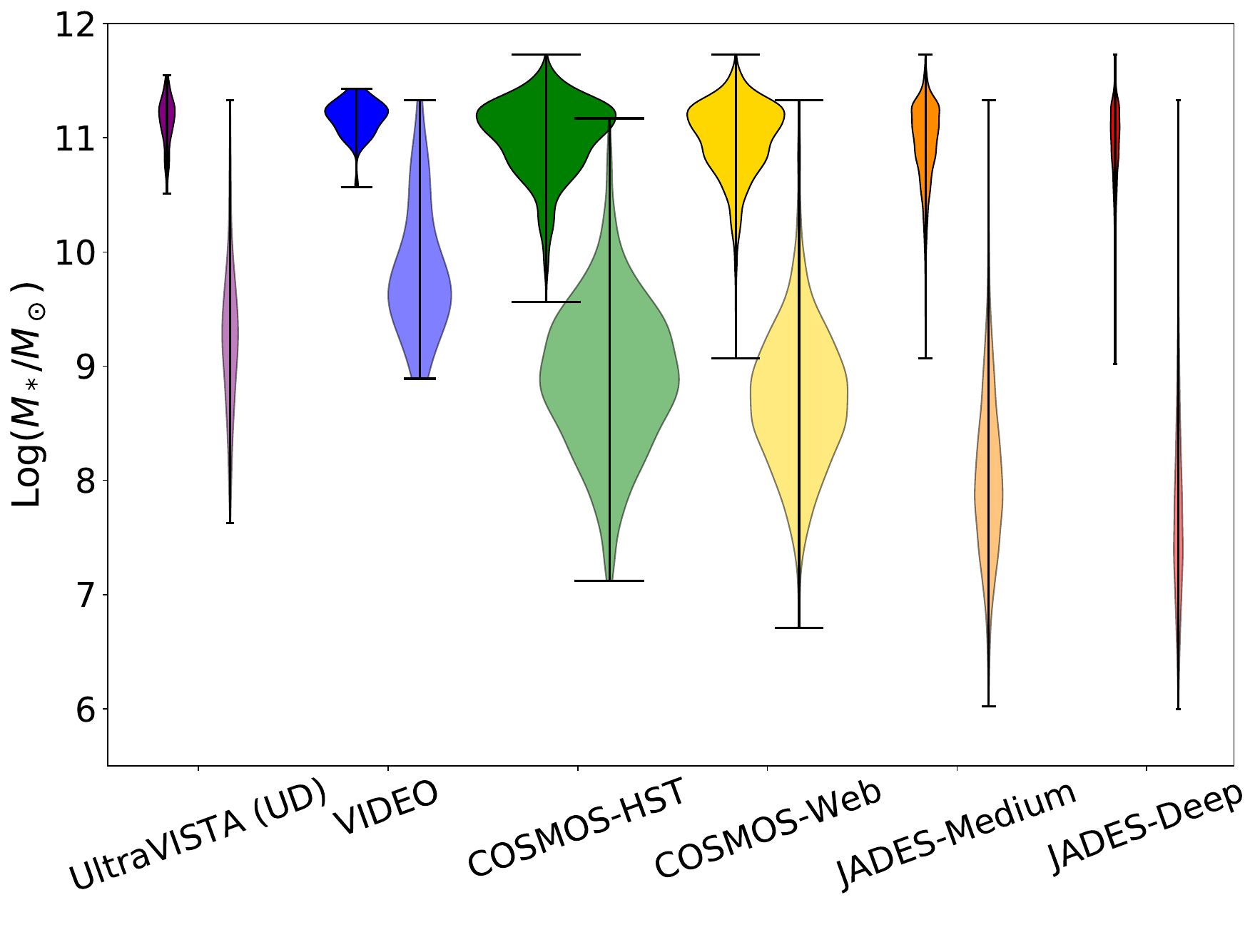}
    \caption{}
    \label{fig:ensemble_boxplots_mass}
\end{subfigure}
\begin{subfigure}{0.33\textwidth}
    \centering
    \includegraphics[width=\textwidth]{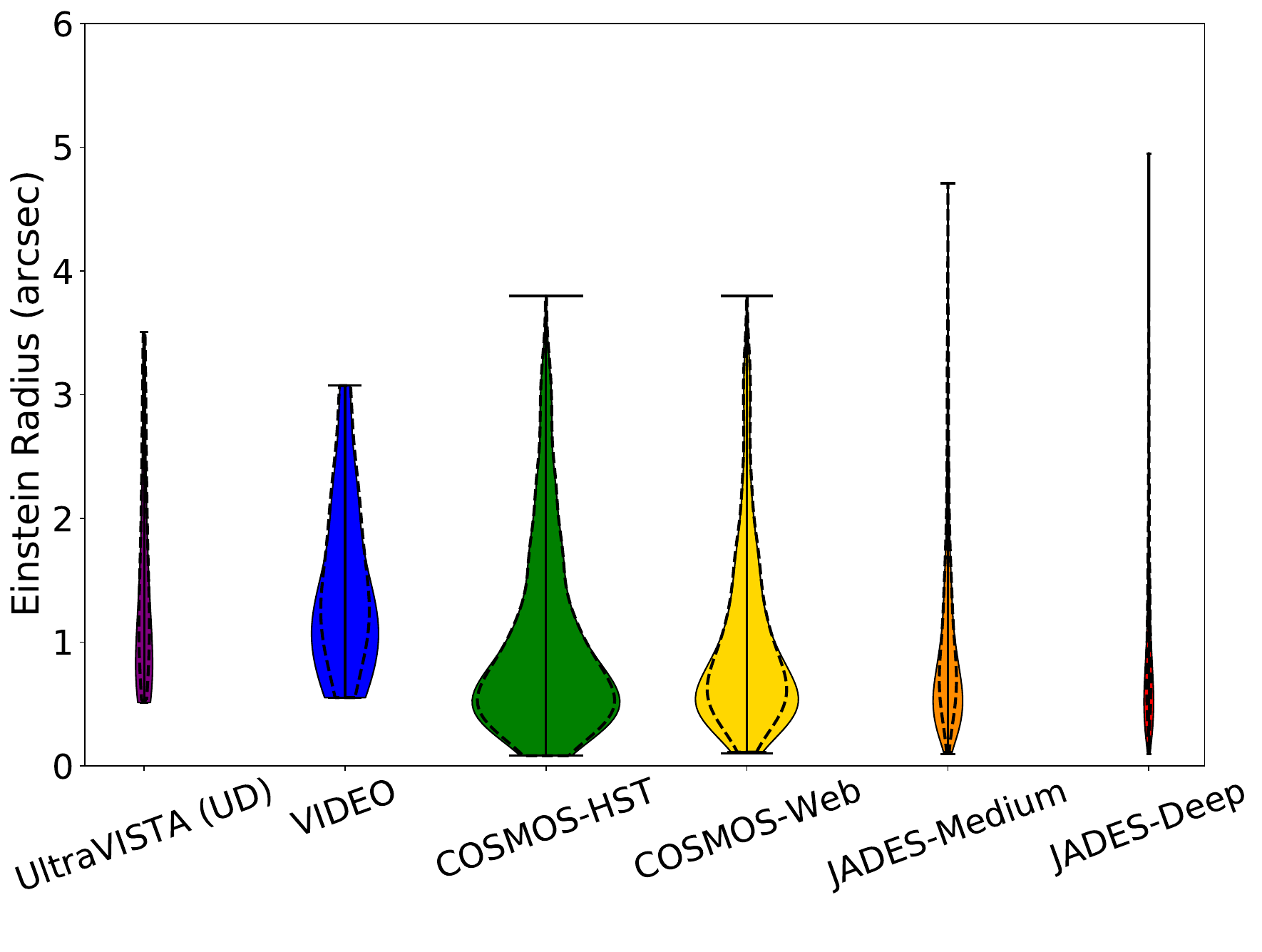}
    \caption{}
    \label{fig:ensemble_boxplots_tE}
\end{subfigure}
\caption{a) Redshift, b) stellar mass and c) Einstein radius distribution of all detectable lenses across all filters in the given programs. The widths of the violin plots are scaled by the expected number of lenses expected in each program. The source galaxy values for redshift and mass are shown by the faded plots. The dashed lines in (c) refer to systems detectable without lens subtraction. Note, the surveys are of significantly different sizes so the extreme values will be very unlikely to be observed, especially in the smaller JADES-Medium/Deep surveys.}
\label{fig:ensemble_boxplots}
\end{figure*}

Most of the detectable lens population in the surveys investigated are located at $z\sim1$ (Fig. \ref{fig:ensemble_boxplots_redshift}) and have high stellar mass ($\sim10^{11}M_\odot$, Fig. \ref{fig:ensemble_boxplots_mass}) although the deeper space-based surveys (\emph{JWST}/\emph{HST}) can detect lower mass lenses. The majority of lenses ($\sim70$ per cent) are quiescent, as flagged by the JAGUAR mock catalogue. The lenses flagged as star-forming are redder ($\Delta (g-r) \sim 1$mag) and much more massive (by a factor of $\sim 10^{4.5}$) than typical SFGs in the simulation at those redshifts. 

\begin{figure*}
\centering
  \includegraphics[width=\textwidth]{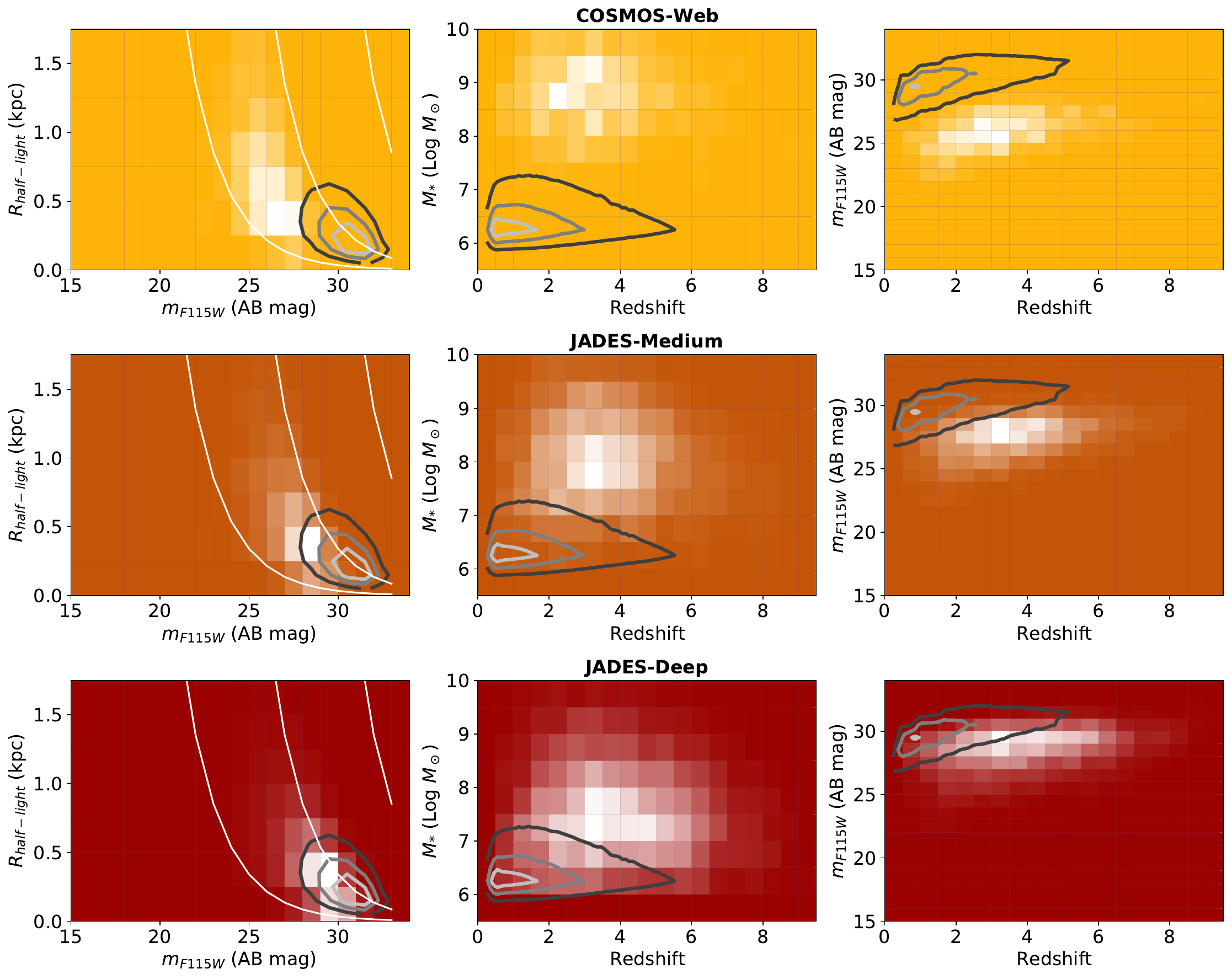}
\caption{Properties of the source population of all detectable systems in the COSMOS-Web, JADES-Medium and JADES-Deep (top-bottom) systems, across all filters. The contours reflect the properties of the whole population in the JAGUAR catalogue, while the 2D histogram displays the detectable sources in each survey. The white lines in the first column show curves of constant surface brightness.}
\label{fig:detectable_source_properties}
\end{figure*}

\begin{figure}
\centering
\includegraphics[height=5.5cm]{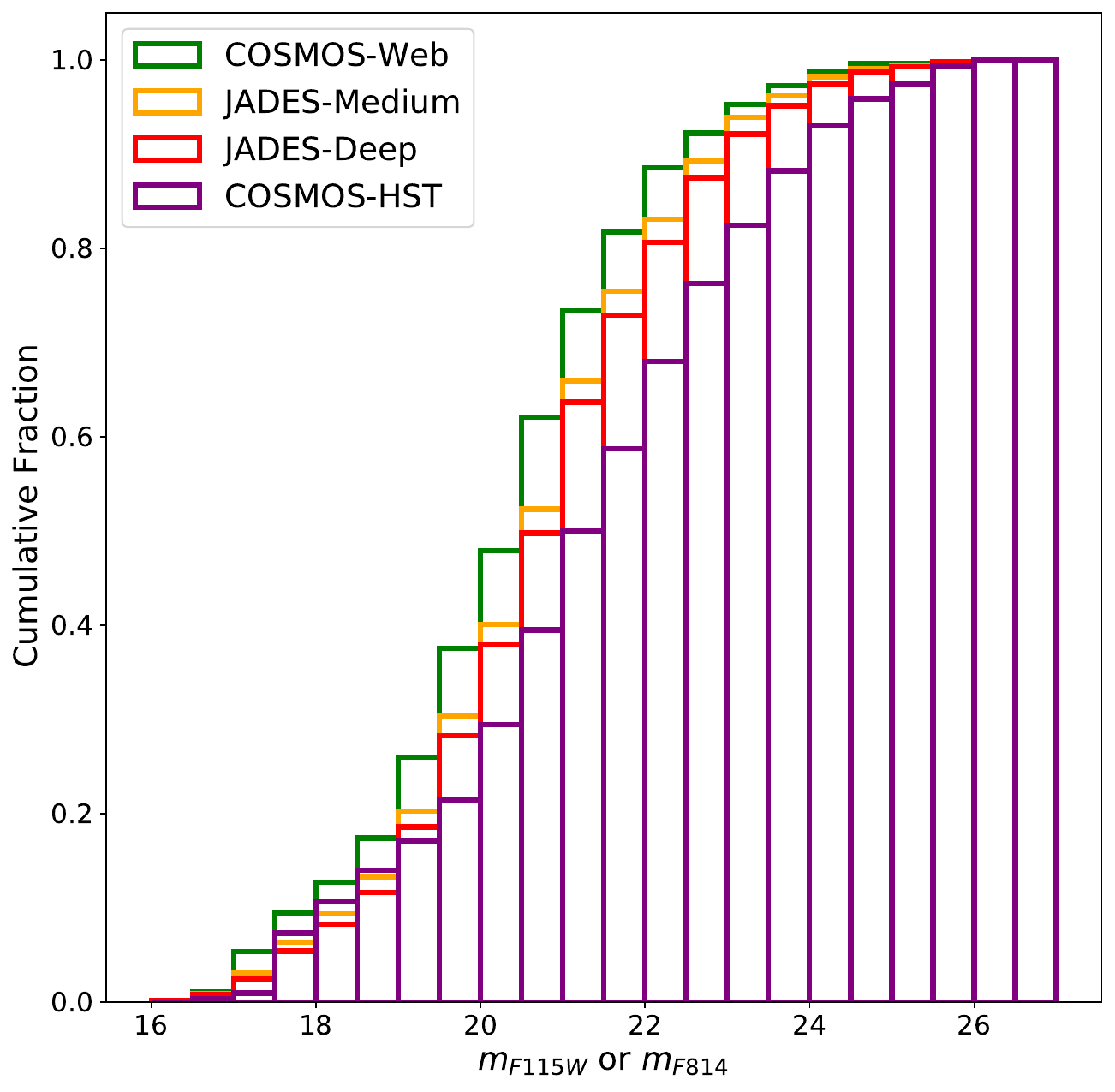}
\caption{
Cumulative histogram of the lens magnitudes of the detectable lenses in COSMOS-HST ($m_{\rm F814W}$ band) and \emph{JWST} programs ($m_{\rm F115W}$ band). Roughly 80 per cent of the lenses would be detected in a search of galaxies with $m_{\rm F115W}$<22 for \emph{JWST} and $m_{\rm F814W}<23$ for \emph{HST}.}
\label{fig:lens_magnitude_jwst_unique}
\end{figure}

The sources predominantly lie on the star-forming main sequence at redshifts z$\sim$2--4 (Fig. \ref{fig:ensemble_boxplots_redshift}) and are intrinsically brighter than the general population (Fig. \ref{fig:detectable_source_properties}). 
The deeper programs target higher redshift sources with lower surface brightness at each redshift. The source galaxies become more typical of the general population with depth; this can be seen in the right-hand column of Fig. \ref{fig:detectable_source_properties} where the modal bins of F115W magnitude with redshift are brighter by $\Delta m_{\rm F115W} \sim 5, 3$ and $1$ than the general population for COSMOS-Web, JADES-Medium and JADES-Deep respectively.\\
Lens searches can be targeted (e.g. \citet{Faure2008}), for example selecting targets based on their magnitude, or untargeted (e.g. \citet{More2016}). Such selections can significantly speed up a lens search but can come at the expense of reduced completeness. From our simulations we can measure the effect of such a selection. Fig. \ref{fig:lens_magnitude_jwst_unique} shows the cumulative fraction of detectable lens systems with F115W magnitude across the \emph{JWST} programs; a simple lens selection of $m_{\rm F115W}\lesssim22$ would identify 80 per cent of the available systems in \emph{JWST}. $96$ per cent of our simulated lens systems detectable in COSMOS had $m_{F814W}<25$mag (i.e. the cut used by \citet{Faure2008} and \citet{Jackson2008}), indicating this cut would only have a minor affect on completeness.\\

\subsubsection{Detection of multiple imaging}\label{multiple imaging}
\begin{figure}
  \centering
  \includegraphics[height=100pt]{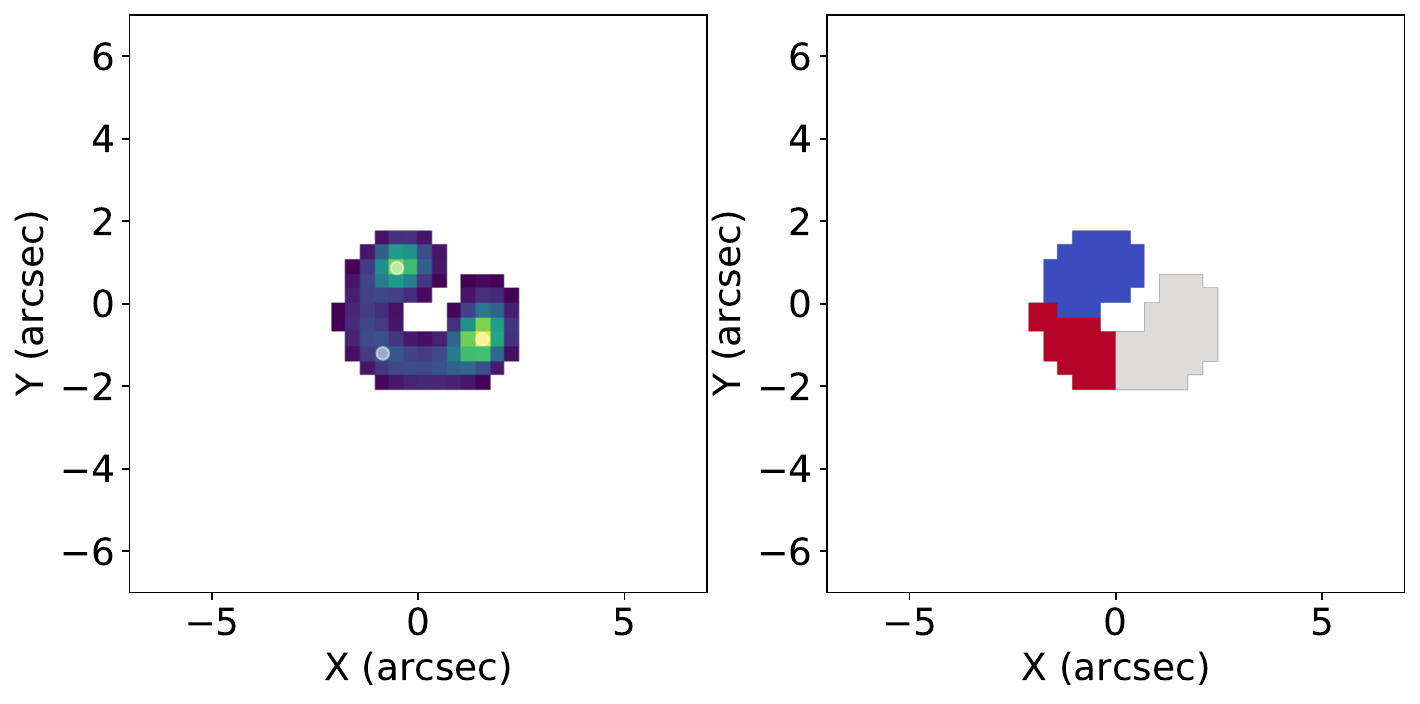}
  \caption{An example of a lensed image (L) along with its labelled sub-images (R). The pixels with S/N$<$1 have been masked out. The white points in the left image indicate the local maxima.}
  \label{sub_image_example}
\end{figure}

From our simulations, we also make an estimate on the frequency of detecting multiple imaging in our mock galaxy-galaxy strong lenses. For clarity, we will here call one of the multiple images from the same source a sub-image, e.g. a single arc in an image containing two arcs from the same source. To split up the lensed image into sub-images we perform gradient ascent from all pixels with S/N $\geq$ 1 (per pixel) to find their associated maxima. We label all pixels which trace to the same maxima as part of the same sub-image. For multiple imaging to be observed, the sub-images must each be detectable and resolvable from each other. To be detectable, we require at least one sub-image to have S/N>20 and any counter-image to have S/N$>$5. To be resolvable we require a sub-image to be positioned further than the FWHM of the PSF (or seeing) from another sub-image's peak. An example of such sub-images is shown in Fig. \ref{sub_image_example}.\\
The proportion of simulated systems which have detectable multiple imaging is $\sim$55--80 per cent with deeper surveys having a larger proportion. If we impose a S/N>20 constraint on all sub-images, this reduces to $\sim$20--50 per cent. These are systems where the multiple images would be easily visible and so would likely be included in the `A-grade' candidates in a visual search. The presence of multiple imaging adds confidence to lens identification and would provide tighter constraints for lens modelling. For a COSMOS-HST search akin to \citet{Faure2008}, we find $\sim14$ of our simulated detectable systems would have detectable multiple images (with a $S/N>5$ limit for counter-images) and $\sim23$ following lens subtraction; this is in agreement with the number of systems identified by \citet{Faure2008} (20, some of which were identified after lens modelling).

\subsection{Extrapolating to wide-area surveys}\label{Extrapolating to wide-area Surveys}
In this section, we explore ball-park estimates for wide-area forthcoming surveys based on our simulations with the caveat that due to the small area of JAGUAR, the most massive galaxies and therefore most likely lenses, are absent from the simulations. We account for this incompleteness by determining the proportion of massive lens galaxies (i.e. the fraction with $M_{*}>10^{11.5}M_{\odot}$) in the published strong lens samples SuGOHI \citep{Sonnenfeld2019} and KiDS \citep{Petrillo2017} with z$>$0.2 to match the lower limit of the JAGUAR catalogue. These suggest that, from the mass distributions alone, our wide-area results would underestimate the true occurrence rate by a factor of 1.3 (KiDS) to 3 (SuGOHI). We regard this as a lower limit since we note that both KiDS and SuGOHI will have their own incompletenesses that are not counted for in the factors above, and if we include z$<$0.2 lenses, we might expect another factor of 2 (based on the fraction of z<0.2, $M_{*}>10^{11.5}M_{\odot}$ lens galaxies in SLACS \citep{Auger2009}). While these estimates are therefore approximate, they remain useful guides (lower estimates) of the number of strong lenses that can be expected in forthcoming NIR surveys.

The Euclid Wide survey will cover an area of $15,000\,\deg^2$. Using LENSPOP, \citet{Collett2015} predicted 170,000 lenses could be detectable in the \emph{Euclid VIS} band. In this study, we predict 95,000 strong lenses in the \emph{VIS} channel, i.e. approximately half of \citet{Collett2015}'s number.  This apparent discrepancy arises from differences in the methods adopted as well as differences in the properties of the basis catalogues. As mentioned above, the JAGUAR catalogue is missing massive galaxies (a comparison of velocity dispersion distributions show that $\sim$30\% of LENSPOP galaxies with the highest velocity dispersions are absent from the JAGUAR catalogues), and JAGUAR has a higher completeness at faint magnitudes than the LSST catalogue; at the faint end the galaxies in JAGUAR outnumber those in the LSST catalogue used in LENSPOP by a factor of $\sim$3. This fainter depth plays a significant role if we were to use a background noise level/image threshold as adopted by \citet{Collett2015} (approx $\sim$2.5 lower than the limit we adopted to define detectablilty) which would result in $\sim$ 400,000 strong lenses being detectable in \emph{Euclid} VIS from our estimates. Given the differences in the methods as well as the complexities of detecting lenses in real on-sky data, the estimates remain broadly consistent. Those determined with this method are conservative due to the high threshold assumed and lack of the most massive galaxies in JAGUAR. LENSPOP provides a mid-range estimate owing to a lower threshold but also potentially underestimates faint sources (as the JADES \citep{Rieke2023b} results would suggest). Therefore a larger number of strong lenses than predicted by LENSPOP may be found in \emph{Euclid}-VIS but they are likely to be at the faint end in source magnitude and may be harder to securely identify as strong lenses. 

Considering the Euclid NISP bands together, for which lens estimates have not yet been published, we anticipate  $\sim70,000$ strong lenses will be detectable (noting again our estimates are conservative). Interestingly we find 40 per cent of these are not detectable in the \emph{VIS} band - we discuss this further in the Section \ref{Comparison of NIR vs visible}. Considering the forthcoming wide-area \emph{Roman} surveys, previous estimates by \citet{WeinerPaper2020} (based on LENSPOP) suggested that $\sim$17,000 strong lenses would be detectable in the J129 band in $2000\,\deg^2$. In contrast, our estimate ($\sim 90,000$) in $1700\,\deg^2$ is significantly higher. This can be attributed to the difference in the zeropoints used (26.4 in this study vs 23.9 in \citet{WeinerPaper2020}). Our zeropoint is consistent with the published transmission curves for \emph{Roman} (Section \ref{SN_Calculation}).

Overall for the survey depths provided in Table \ref{tab:N_occurence_table_p3}, our simulations suggest that we can expect to detect $\mathcal{O}(10^5)$ lenses in the \emph{Euclid} and \emph{Roman} surveys. More precise estimates, utilising a wide-area simulation, are beyond the scope of this paper.

\section{Discussion}\label{Discussion}
We now discuss the results of our simulations. We first focus on the number of detectable lens systems in forthcoming \emph{JWST} surveys, before generalising to the other surveys investigated in this study. We then explore the properties of such detectable systems, and how they vary with wavelength.

\subsection{Number density of detectable strong lenses systems}
\subsubsection{Prospects for JWST surveys}
Our simulated catalogue is best suited for small-area, deep surveys such as those of \emph{JWST}. Specifically, we generate lensing predictions for COSMOS-Web, JADES-Medium and JADES-Deep that span a wide range of depths and areas and thus provide indicative results for typical \emph{JWST} surveys, present or future. Of the three surveys considered here, our simulations suggest the COSMOS-Web survey \citep{Casey2022} will contain the most strong lens systems ($\sim 65$ systems across all filters). This is broadly consistent with the preliminary estimates from \citet{Casey2022} which predicted O(100) lenses (however since we apply detectability constraints it is unsurprising the number we predict is slightly lower).  We find the JADES-Medium and Deep programs include $\sim 25$ and $\sim 10$ detectable systems, respectively. \\
A visual inspection of preliminary COSMOS-Web data\footnote{\url{https://cosmos.astro.caltech.edu/page/cosmosweb}} totalling $77\,arcmin^2$, reveals one high-grade galaxy-galaxy lens (previously identified in \emph{HST} data by \citet{Pourrahmani2018}), a group-scale lens (which would not be found in our simulations), and a lower grade candidate. From our simulations, we expect $\sim2$ lenses in such a field which is consistent with these preliminary findings.\\
Even though NIRCam can only produce pencil-beam surveys, the cumulative area observed across all the surveys that it will take will inevitably lead to serendipitous lens discovery. Indeed, the total area of \emph{JWST} surveys for Early Release Science and in Cycle 1 is $>1\,\deg^2$ \citep{Windhorst2023}, double that of COSMOS-Web alone. Thus, we predict that the number of galaxy-galaxy lenses identifiable in early \emph{JWST} fields will likely surpass 100, a non-negligible fraction of the total number of lenses currently known, with corresponding excellent image quality and photometry in the near-and mid-infrared. A future archival search akin to \citet{Pawase2014} would capture the full range of strong lenses in the \emph{JWST} survey fields.

\subsubsection{Prospects for VIDEO, UltraVISTA, Euclid \& Roman}
\begin{figure}
\centering
\includegraphics[width=\columnwidth]{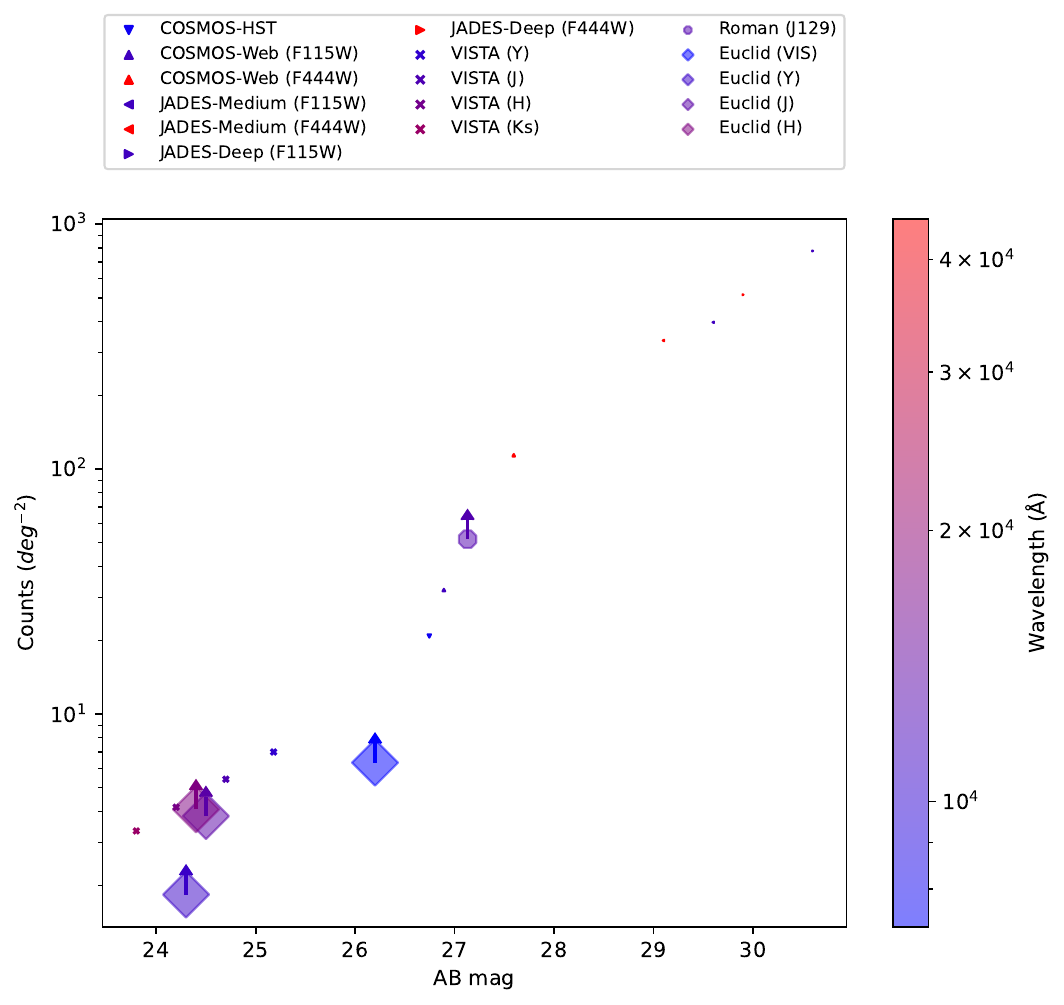}
\caption{A comparison of the number density (lens systems per $\deg^{2}$) of lenses detectable in a range of existing and future surveys. The area of the data-points scales with the (square root of the) survey area, and are coloured by their central wavelength.}
\label{fig:prospects for surveys}
\end{figure}

Strong lensing science will change substantially with the arrival of wide-area surveys such as \emph{Roman} and \emph{Euclid} identifying tens of thousands of lenses. This will allow the lens population to be treated in a statistical fashion to determine the properties of the strong lens population as a whole (e.g. constraining the typical dark matter profile, \citet{Sonnenfeld2021}). Fig. \ref{fig:prospects for surveys} shows the number density of lens systems detectable for a range of existing and future surveys. Although the \emph{JWST} programs mark a step-change in survey depth at NIR/MIR wavelengths and the number density of detectable lenses is an order of magnitude or so greater than that expected for \emph{Euclid VIS}, their survey areas are $\mathcal{O}(10^5)$ smaller. Regarding number density however, these small, deep surveys surpass the wide area surveys we consider. For example the number of detectable lenses in COSMOS-Web is expected to be similar to that in the VIDEO survey and greater than that in UltraVISTA. The smaller, higher resolution programs in \emph{JWST} will produce small numbers of high image-quality (PSF$\sim0.05\arcsec$) lenses for precision modelling of individual sources. Naturally, the numbers discussed here are merely a guide of what to expect, but what fraction of these are scientifically useful is dependent on the particular astrophysical or cosmological goals to be investigated. 
\subsubsection{Variation in the number of detectable lenses with waveband}
The band-pass used for the lens search also influences the number of detectable lenses. This varies due to the difference in survey depth between bands. For COSMOS-Web, the F444W band has $\sim3.5\times$ more detectable systems than the F115W band. 
In the VIDEO survey, the greater depth of the \emph{Y} band reveals $\sim 2.1\times$ more detectable lenses compared to the \emph{Ks} band. Using multi-band imaging when conducting a lens search improves the ability of lens detection methods to separate the lens and source galaxies (e.g. \citet{Metcalf2019}). Furthermore, more precise photometry in the NIR will give more accurate derived properties such as photometric redshifts and stellar masses.

\subsection{Properties of the detected strong lenses}
Beyond simply the number of lenses detectable in different surveys, with the SED and photometry information present in our parent catalogue, we explore how the properties of these lens systems also vary. 
\subsubsection{Variation of $z_L$, $M_{*}$ and $\theta_E$ with survey}
Fig. \ref{fig:ensemble_boxplots} shows the properties of the detectable lens population for a range of surveys across their respective filters. The lens stellar-masses and redshifts are broadly unchanged by survey, all clustered around $10^{11}M_\odot$ and $z_{L}\sim1$. Unsurprisingly, the space-based surveys (\emph{HST} and \emph{JWST}) are found to probe smaller $\theta_E$ values than the ground-based VIDEO and UltraVISTA programs. This is due to the smaller PSF and pixel scale of \emph{JWST} and \emph{HST} compared to the much larger seeing of the ground-based surveys. \\
Comparing the lens systems detectable in the three COSMOS field surveys (COSMOS-Web, COSMOS-HST and UltraVISTA), the smaller angular resolution of \emph{HST} and \emph{JWST} allows these telescopes to identify a larger number of systems, extending to lower lens masses ($\sim10^{10.5}M_{\odot}$) though the majority of lenses in all cases still have masses $\sim10^{11}M_{\odot}$. The benefit of utilising lens-subtracted images also varies dependent on the survey as shown by Fig. \ref{fig:ensemble_boxplots_tE}. Lens subtraction has the most use for the ground-based VIDEO search (an average gain of 50 per cent) as the lenses tend to be brighter and thus more easily able to shroud any lensed images.

\subsubsection{Prospects for detecting lensed high-z sources with JWST}
One of the primary goals of the \emph{JWST} blank field surveys is to detect galaxies at the very highest redshifts. While many such galaxies have been located in cluster fields (z$\sim$10, \citet{Zitrin2014,Coe2013}), here we look at high-redshift (z$>$6) galaxy-galaxy lenses which are present in our simulations and could be detected in blank field surveys. Using our simulations we can determine the probability of detecting galaxy-galaxy lenses at high redshift. Drawing $N$ source redshift values randomly 100,000 times from the simulated detectable source redshift distribution (where $N$ is the number of detectable lenses in the survey from Table \ref{tab:N_occurence_table_p2}), we find a $90$ per cent likelihood of a z$>$6 lensed object in COSMOS-Web. Such sources are more massive and luminous ($M_*\sim 10^{8.5}, m_{\rm F115W}\sim 26.5$mag) than typical high-redshift galaxies, as shown in Fig. \ref{fig:detectable_source_properties}. Although the source-redshift distributions of the deeper JADES programs also extend to high redshift and contain more typical high-z sources ($M_*\sim 10^{7.5}, m_{\rm F115W}\sim 29$mag), their small survey areas mean few if any of these would likely be observed.

\subsubsection{Comparing detected lens systems in the NIR verses the visible}\label{Comparison of NIR vs visible}
Based on our simulations with full visible-to-NIR SEDs, we can explore whether strong lens searches in the NIR bring additional information over the traditional searches historically conducted in the visible (e.g. \citet{Sonnenfeld2020,Jacobs2019}). First we consider the narrow-field JWST surveys compared to HST. Owing to the greater depth and wider wavelength coverage, nearly all those detectable with \emph{HST} would be detected by \emph{JWST} while only $\sim40$ per cent of our simulated systems detectable in COSMOS-Web would be also detectable in COSMOS-HST. Therefore, \emph{JWST} can provide verification (and/or rejection) of candidate lenses previously discovered in the COSMOS-HST field. The sources absent from \emph{HST} but detectable by COSMOS-Web are typically fainter by $\Delta m_{\rm F814}=1.3$ (in lensed magnitude, where $\Delta$ denotes the difference in medians) and higher redshift ($\Delta z_{S}\sim0.3$). Thus the COSMOS-Web strong lens samples will likely contain higher redshift lensed sources than are seen in the \emph{HST} searches.\\
In contrast to \emph{JWST} vs \emph{HST}, the \emph{Euclid VIS} band is substantially deeper and has better image quality than its NISP channels which increases the number of lenses detectable in the visible compared to the NIR. Despite the lower image quality, the NISP bands would add colour information which would make strong lens identification easier than using the \emph{VIS} band alone. This would be true for even those \emph{VIS}-identified strong lenses that lie below our SNR$>$20 threshold for detectability in the NISP channels. The NISP bands will offer new systems too. Across all the lens systems detectable with \emph{Euclid}, 43 per cent are only detectable in the \emph{VIS} channel, 35 per cent in both \emph{VIS} and NISP channels and 22 per cent are detectable only in the NISP channels. The lensed sources not detectable in the \emph{VIS} channel are typically at lower redshifts ($\Delta z_{S} \sim 0.8$) and have redder colours ($\Delta(\emph{VIS}-Y)\sim0.9$). On average, they are also dustier (the \emph{V}-band dust attenuation differing by $\Delta \tau_V \sim 0.3$) and have older stellar populations ($\Delta Log(t_{\rm max}/yr) \sim 0.5$) where $t_{\rm max}$ is the maximum age of stars in the galaxy. This suggests that (a) it is worthwhile running lens searches utilising the \emph{Euclid} NISP bands as well as \emph{Euclid VIS} and (b) such a search would include a component of dusty, moderate redshift lensed sources (e.g., \citet{Geach2015}). Lensed,  ultracompact quiescent galaxies (e.g. \citet{Muzzin2012}) may also be detected. While bright in $K$, the frequency of z$>$2 lensed quiescent galaxies is estimated to be 0.5-1$\,\deg^{-2}$ \citep{Muzzin2012} so would be unlikely to be detected in \emph{JWST} surveys of similar size to COSMOS-Web, but highly likely to be observed with \emph{Roman} and \emph{Euclid}. 

Beyond purely increased number, our predictions show that we can expect interesting lens and source science to be addressed through new samples of lenses emerging from ongoing and forthcoming ground- and space-based visible and NIR surveys.

\section{Conclusions}\label{Conclusion}
We generate lensing frequency estimates for strong lenses detectable in existing and forthcoming telescope surveys. To do this we use the JAGUAR galaxy catalogue, combined with the DREaM catalogue to provide dark matter profiles, to generate a realistic galaxy population from which to draw lens and source galaxies. The resultant galaxy catalogue is small in area and suitable for studying the lens population detectable in $\lesssim10deg^2$ surveys such as \emph{JWST},VIDEO and UltraVISTA. We explore the number and properties of strong lenses in surveys utilising \emph{JWST}, \emph{HST}, VISTA, \emph{Euclid} and \emph{Roman} telescopes. Our conclusions are as follows:
\begin{itemize}
    \item \emph{JWST} will likely identify $\gtrsim$100 lenses from the \emph{JWST} Early Release Science and Cycle 1 data. Out of the \emph{JWST} programs investigated, chosen to be illustrative of the range of surveys \emph{JWST} will undertake, COSMOS-Web will deliver the largest number of detectable lenses ($\sim 65$) across all bands, with a further $\sim25$ across JADES-Medium and JADES-Deep. 
    \item Out of the surveys we investigate, our predictions suggest that multiple imaging could be detectable (at SNR$>$5) in 55-80 per cent of the detectable strong lens sytems. Deeper surveys have higher proportions of lensed systems with detectable multiple images.
    \item Our simulations confirm that lens subtraction has the most benefit for ground-based surveys, e.g., increasing the yield by $\sim 50$ per cent for the VIDEO survey.
    \item \emph{JWST} will be able to detect fainter, higher redshift sources than those identified by \emph{HST}, extending to $z_{S}>6$.
    \item Of all the strong lens systems detectable by \emph{Euclid}, $\sim20$ per cent would only be detected by a lens search which included the NISP channels. These could include dusty $z\sim2$ lensed galaxies which would be missed in an optical search.
    \item Wide-area NIR telescopes (\emph{Euclid} and \emph{Roman}) combined with the deep JWST surveys will provide a step-change in the number and diversity of strong lenses, enabling a wide range of science goals to be addressed.
\end{itemize}

\section*{Acknowledgements}

We thank the referee, Tom Collett, for his useful and insightful comments. This work has relied on the DREaM and JAGUAR galaxy catalogues and we thank \citet{Drakos2022} and \citet{Williams2018} for these data.\\
PH would like to thank Matt Jarvis for very useful discussions during this work and Lance Miller, Mischa Schirmer and Harry Ferguson for information regarding the \emph{Euclid} and \emph{Roman} surveys. 
PH acknowledges funding from the Science and Technology
Facilities Council, Grant Code ST/W507726/1. The UltraVISTA and VIDEO data are based on data products from observations made with ESO Telescopes at the La Silla Paranal Observatory under ESO programme ID 179.A-2005 and ID 179.A-2006 and on data products produced by CALET and the Cambridge Astronomy Survey Unit on behalf of the UltraVISTA and VIDEO consortia.
This work made use of Astropy:\footnote{http://www.astropy.org} a community-developed core Python package and an ecosystem of tools and resources for astronomy \citep{Astropy2022}. It has also made use of the \texttt{numpy} \citep{Numpy2020}, \texttt{scipy} \citep{Scipy2020}, and \texttt{matplotlib} \citep{Matplotlib2007} Python packages.

\section*{Data Availability}
The main data sets underlying this article are available at \href{https://fenrir.as.arizona.edu/jaguar/}{https://fenrir.as.arizona.edu/jaguar/} and \href{https://www.nicoledrakos.com/dream}{https://www.nicoledrakos.com/dream}.  The adapted catalogues, along with estimates for further NIR surveys will be shared upon request to the corresponding author.



\bibliographystyle{mnras}
\bibliography{bibliography} 





\bsp	
\label{lastpage}
\end{document}